\documentclass[a4paper,fleqn,review]{cas-sc}

\usepackage[authoryear]{natbib}
\usepackage[amsstyle]{cases}
\usepackage{makecell}
\usepackage{hyperref}
\hypersetup{
citebordercolor=white,
linkbordercolor=white,
urlbordercolor=blue}
\newcommand{\al}{$\rm ^{26}Al$ }
\newcommand{\fe}{$\rm ^{60}Fe$ }
\newcommand{\Xs}{$\rm X_{S,0}$ }
\newcommand{\feratio}{$\rm ^{60}Fe/^{56}Fe$ }
\newcommand{\rcmf}{$\phi_{C}$ }
\newcommand{\etar}{$\eta_0$ }
\newcommand{\Rem}{$Re_m$ }
\newcommand{\fcmb}{$F_{\rm CMB}$ }

\newcommand{\fep}{$\rm ^{60}Fe$}
\newcommand{\Xsp}{$\rm X_{S,0}$}
\newcommand{\feratiop}{$\rm ^{60}Fe/^{56}Fe$}
\newcommand{\rcmfp}{$\phi_{C}$}
\newcommand{\etarp}{$\eta_0$}
\newcommand{\Remp}{$Re_m$}
\newcommand{\fcmbp}{$F_{\rm CMB}$}

\begin{document}
\let\WriteBookmarks\relax
\def\floatpagepagefraction{1}
\def\textpagefraction{.001}

\shorttitle{Early and elongated epochs of planetesimal dynamo generation}

\shortauthors{Sanderson et~al.}

\title [mode = title]{Early and elongated epochs of planetesimal dynamo generation}                      



%
\author[1]{Hannah R. Sanderson}[orcid = 0000-0001-5842-6985]

\cormark[1]

\fnmark[1]

\ead{hannah.sanderson@earth.ox.ac.uk}

\credit{Conceptualization, Methodology, Software, Writing - Original Draft}

\affiliation[1]{organization={Department of Earth Sciences},
    addressline={University of Oxford, South Parks Road}, 
    city={Oxford},
    postcode={OX1 3AN}, 
    country={UK}}

\author[1]{James F.J. Bryson}[orcid = 0000-0002-5675-8545]
\credit{Conceptualization, Writing - Review \& Editing, Supervision}

\author[1]{Claire I.O. Nichols}[orcid = 0000-0003-2947-5694]

\credit{Conceptualization, Writing - Review \& Editing, Supervision}

\cortext[cor1]{Corresponding author}


\begin{abstract}
Accreting in the first few million years (Ma) of the Solar System, planetesimals record conditions in the protoplanetary disc and are the remnants of planetary formation processes. The meteorite paleomagnetic record carries key insights into the thermal history of planetesimals and their extent of differentiation. The current paradigm splits the meteorite paleomagnetic record into three magnetic field generation epochs: an early nebula field ($\lesssim$5\,Ma after CAI formation), followed by thermal dynamos ($\sim$5--34\,Ma after CAI formation), then a gap in dynamo generation, before the onset of core solidification and compositional dynamos. These epochs have been defined using current thermal evolution and dynamo generation models of planetesimals. Here, we demonstrate these epochs are not as distinct as previously thought based on refined thermal evolution models that include more realistic parametrisations for mantle convection, non-eutectic core solidification, and radiogenic \fe in the core. We find thermal dynamos can start earlier and last longer. Inclusion of appreciable \fe in the core brings forward the onset of dynamo generation to $\sim$1--2\,Ma after CAI formation, which overlaps with the existence of the nebula field. The second epoch of dynamo generation begins prior to the onset of core solidification  \- this epoch is not purely compositionally driven. Planetesimal radius is the dominant control on the strength and duration of dynamo generation, and the choice of reference viscosity can widen the gap between epochs of dynamo generation from 0--200\,Ma. Overall, variations in planetesimal properties lead to more variable timings of different planetesimal magnetic field generation mechanisms than previously thought. This alters the information we can glean from the meteorite paleomagnetic record about the early Solar System. Evidence for the nebula field requires more careful interpretation, and late paleomagnetic remanences, for example in the pallasites, may not be evidence for planetesimal core solidification.  

\end{abstract}



\begin{highlights}
\item Planetesimal thermal dynamos can begin before the dissipation of the nebula field.  
\item Core solidification is not required to trigger a second epoch of dynamo generation.
\item Planetesimal dynamos can last several 100 Ma longer than previously thought.
\item Planetesimal radius and reference viscosity have a strong effect on dynamo duration.
\end{highlights}


\maketitle

\section{Introduction} 
 
The strength and duration of magnetic fields generated within planetesimals can provide unique insight into the differentiation and thermal histories of rocky bodies in the early Solar System. For a rocky body to internally generate a magnetic field (a dynamo), it must have a partially molten metallic core in which flow is being driven, for example by convection. Magnetic remanences recorded by meteorites indicate that some planetesimals met these criteria during their thermal history, providing evidence for planetesimal differentiation and vigorous core convection. 

Three strands of investigation can be combined to allow observations of magnetic field generation in planetesimals to constrain planetesimal behaviour. Firstly, magnetic remanences measured from meteorites have provided evidence for dynamos on specific parent bodies and radiometric dating of these meteorites provide estimates for the age of dynamo activity \citep[e.g.][]{maurel_meteorite_2020}. Secondly, thermal evolution models can be used to explore parent body properties, such as radius and accretion time, for which the requirements for magnetic field generation are met \citep[e.g.][]{nimmo_energetics_2009,scheinberg_core_2016,dodds_thermal_2021}. For undated magnetic remanences, thermal evolution models can provide possible dynamo timing constraints, for example based on the timing of core crystallisation \citep[e.g.][]{tarduno_evidence_2012,bryson_long-lived_2015,maurel_meteorite_2020,nichols_time-resolved_2021}. Finally, thermal evolution models can also predict whether a remanence is likely to have a dynamo or solar nebula origin \citep[e.g.][]{SterenborgCrowley2013,bryson_constraints_2019,dodds_thermal_2021}. Thermal evolution models to-date include incomplete or simplified physical descriptions of planetesimals. Here, we use refined thermal models to re-evaluate the meteorite paleomagnetic record. 

\subsection{Meteorite paleomagnetic record}
\begin{figure*}
    \centering
    \includegraphics[width=1\textwidth]{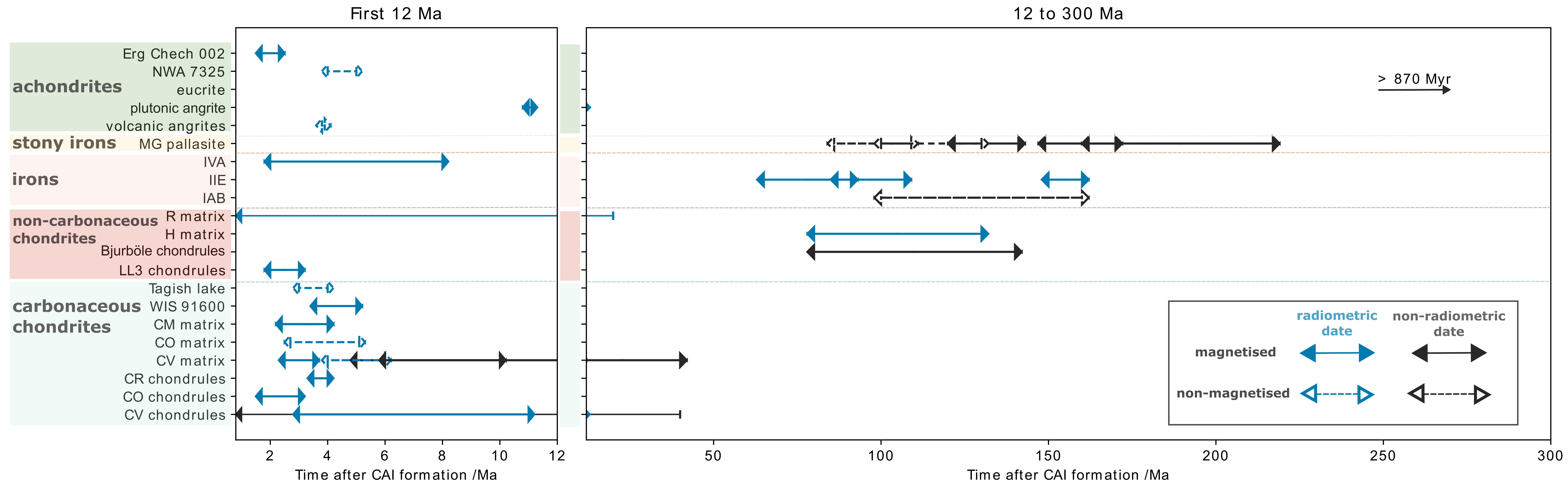}
    \caption{The meteorite paleomagnetic record. Blue and red boxes indicate undifferentiated (chondritic) meteorite groups and green, yellow and pink boxes indicate differentiated (achondritic) meteorite groups. Blue and black arrows represent magnetic remanences dated using radiometric methods and thermal evolution models, respectively. Solid arrows with filled caps indicate magnetised samples and dashed arrows with unfilled caps indicate unmagnetised samples. For the angrites, the time range is so short that only the arrowheads are visible. Multiple measurements for the same meteorite group are combined, but measurements of bulk magnetisation vs chondrules are kept separate to distinguish between pre-accretionary and post-accretionary remanences. Unlike the other chondrules, the CV chondrules remanence is not pre-accretionary, because the CV chondrules have experienced aqueous alteration, thermal metamorphism, and possibly brecciation \citep{fu_no_2014,shah_long-lived_2017}. The data table shown in this plot and references for each measurement can be found in the Supplementary Materials.}
    \label{fig:paleomag}
\end{figure*}
Meteorites with similar isotopic compositions are assumed to come from the same parent body and form a group. Paleomagnetic measurements have been recovered from both differentiated (achondritic) and undifferentiated (chondritic) meteorites from both the outer (carbonaceous) and inner (non-carbonaceous) Solar System (Figure \ref{fig:paleomag}).

Three possible sources have been proposed for these magnetisations: the nebula field in the protoplanetary disc, which existed prior to $\sim$4--5\,Ma after calcium-aluminium-rich inclusion (CAI) formation \citep[e.g.][]{fu_no_2014,wang_lifetime_2017,fu_weak_2020,borlina_paleomagnetic_2021,bryson_unified_2023,maurel_4565-my-old_2024}; internally generated dynamos \citep[e.g.][]{tarduno_evidence_2012,wang_lifetime_2017,bryson_paleomagnetic_2019}; and the solar wind \citep{tarduno_magnetization_2017,obrien_arrival_2020}. Magnetisation from the solar wind has been proposed to be three to four orders of magnitude too weak to explain meteorite paleomagnetic remanences \citep{oran_were_2018}. Remanences from the nebula field can be identified by randomly oriented chondrule magnetisation directions \citep[e.g.][]{fu_weak_2020}. 
 
Paleomagnetic remanences can be dated radiometrically if the closure temperature of the radiometric system is close to the blocking temperature of the magnetised mineral. Remanences dated to $<4$\,Ma after CAI formation have been interpreted to have a nebula origin due to the delay in the onset of dynamo generation predicted by previous dynamo models \citep{dodds_thermal_2021}. Meanwhile, remanences that were acquired after the dissipation of the nebula field, $\sim$4--5\,Ma after CAI formation \citep{weiss_history_2021}, are thought to originate from planetesimal dynamos. For systems where radiometric dating is not possible or extrapolation from a dated closure temperature to the Curie temperature is required, thermal evolution models can be used to predict the timing of dynamo generation. For example, thermal evolution models were used to estimate an age for the Main Group pallasite remanant magnetisation based on the timing of core crystallisation \citep{tarduno_evidence_2012}.

\subsection{Thermal evolution models}\label{intro-therm}
Two mechanisms of dynamo generation are considered by thermal models: thermal dynamos, where temperature-induced density differences drive core convection; and compositional dynamos, where density differences between solidifying pure iron and the liquid core enriched in incompatible light elements drive convection. For thermal dynamos, the superadiabatic portion of the core-mantle boundary (CMB) heat flux determines the timing and strength of magnetic field generation. Modelling compositional dynamos requires assumptions regarding the mechanism of planetesimal core solidification. The lower pressures at their CMB compared to planets ($\leq$0.5\,GPa compared to $\geq$8\,GPa) mean planetesimal cores solidify inwards rather than nucleation occurring at the centre of the core \citep{williams_bottom-up_2009}. Several mechanisms for inward core solidification have been proposed, including iron snow where solid iron particles form at the CMB and then fall to the centre of the core triggering convection \citep{ruckriemen_fe_2015,scheinberg_core_2016} and delamination of large, solid, iron diapirs from the CMB \citep{neufeld_top-down_2019}. Due to the complexity of these mechanisms, previous thermal evolution models of mantled planetesimals either ended after the cessation of mantle convection but prior to core solidification \citep{SterenborgCrowley2013,dodds_thermal_2021} or used the duration of eutectic core solidification as a first estimate of the timings of compositional dynamo duration \citep{bryson_constraints_2019,nichols_time-resolved_2021}. Models of outward non-eutectic core solidification have been attempted but either did not model the full history of the body, or only considered cooling by mantle conduction \citep{nimmo_energetics_2009,nichols_time-resolved_2021}. 

Previous thermal evolution models \citep[e.g.][]{SterenborgCrowley2013,bryson_constraints_2019,dodds_thermal_2021} that successfully generate dynamos predict several events in the thermal evolution: \begin{enumerate}[(i)] 
\item Following instantaneous accretion or during gradual accretion, planetesimals are heated by \al decay, causing the body to differentiate into a metallic core and silicate mantle. 
\item Following differentiation, \al preferentially partitions into the silicate mantle. After further heating, the melt fraction in the mantle is high enough to initiate mantle convection. The core is heated by the mantle from above and develops a stable thermal stratification that prevents the onset of a dynamo. 
\item Once the mantle cools below the temperature at the CMB, the core thermal stratification is removed and the CMB heat flux becomes superadiabatic. Core convection triggers the onset of thermal dynamo generation.
\item As the mantle cools, it becomes more viscous until mantle convection ceases. The cessation of mantle convection leads to a sharp decrease in CMB heat flux and core convection drops below the critical value, terminating the dynamo.
\item Eventually the core cools below its solidus temperature and begins to solidify, which can lead to a compositional dynamo and a second epoch of dynamo generation. 
\item Once the core has fully solidified, dynamo generation is no longer possible. 
\end{enumerate} These previous models predict two epochs of dynamo generation: an early thermal dynamo,and a second later compositional dynamo during core solidification. This has led to paleomagnetic remanences $>$65\,Ma after CAI formation being interpreted as evidence of a solidifying core \citep{bryson_long-lived_2015,nichols_pallasite_2016,nichols_time-resolved_2021,maurel_long-lived_2021} and remanences $<$4\,Ma after CAI formation being interpreted as a record of a nebula field due to the delay in start time of thermal dynamos caused by core thermal stratification \citep{bryson_constraints_2019,bryson_constraints_2020,fu_fine-scale_2021}.

The assumptions made in these earlier models are re-explored here. First, previous models neglect heating from \fe in the core, because for low values of primordial \fep, such that \feratiop=$10^{-8}$ \citep{tang_abundance_2012}, only a very small difference in core temperature ($\sim$10\,K) results over the lifetime of the planetesimal \citep{henke_thermal_2013}. However, measurements of the primordial \feratio vary over two orders of magnitude \citep{kodolanyi_iron-60_2022} with an upper estimate of $6\pm2\times10^{-7}$ in the formation region of carbonaceous chondrites \citep{cook_iron_2021}. Therefore, the role \fe on planetesimal core thermal evolution and dynamo generation should be explored explicitly for a range of values.

Second, previous thermal evolution models do not fully combine thermal and compositional dynamo generation, mantle convection, and non-eutectic core solidification. This limits the ability of existing thermal evolution models to predict the full magnetic history of a body.

Third, the effect of mantle viscosity on magnetic field generation has not been investigated in previous thermal models. CMB heat flux is limited by the cooling rate of the mantle above and the thickness of the CMB boundary layer, which are both governed by the mantle viscosity. Therefore, the parameterisation of mantle viscosity as a function of temperature has a critical role in planetesimal thermal evolution. 

\citet{sanderson_unlocking_2025} present a new, refined model, that accounts for all the points discussed above. Here, we apply this new model across the full range of realistic parameters to reassess the existing meteorite paleomagnetic record.

\section{Methods}\label{methods} 
This study uses the 1D spherically symmetric planetesimal thermal evolution and dynamo generation model from \citet{sanderson_unlocking_2025} to predict the timings of dynamo generation in planetesimals with a radius up to 500\,km. This model builds on the previous planetesimal thermal evolution models of \citet{SterenborgCrowley2013}, \citet{bryson_constraints_2019} and \citet{dodds_thermal_2021} by including radiogenic heating from \fe in the core, an adjustable mantle viscosity law, and updated stagnant lid and CMB boundary layer parametrisations that better reflect mantle conditions. The model includes both convection and conduction in the mantle and models magnetic field generation by non-eutectic core solidification. The details and complete mathematical description of this model can be found in \citet{sanderson_unlocking_2025}.

\subsection{Thermal evolution model} 

All model runs begin at 0.8\,Ma after CAI formation with instantaneous accretion of the planetesimal at 200\,K. The body heats up due to radiogenic heating of \al and the Fe-FeS and silicate phases melt until the silicate reaches its critical melt fraction, \rcmfp. At this point, differentiation is assumed to take place instantaneously, as the molten, dense Fe-FeS sinks through the rheologically weak silicates to the centre. The model then consists of two coupled reservoirs: the Fe-FeS core and the silicate mantle. Sulfur is the only light element considered in the core. The surface temperature is held constant at 200\,K throughout the simulation.

In the mantle, heat is transported either by conduction or convection. Mantle convection is assumed to occur in the stagnant lid regime with an isothermal, convective layer between conductive boundary layers at the CMB and the surface. These boundary layers thicken as the mantle cools and convection ceases once the combined CMB and surface boundary layer thicknesses reach the total mantle thickness. The crust is part of the conductive boundary layer at the surface and is not modelled explicitly \citep{sanderson_unlocking_2025}. As a result, our model assumes crustal heat transport is purely conductive and neglects advective heat transport, e.g. heat pipes \citep{moore_heat-pipe_2013}.
The temperature at the CMB is calculated by assuming continuity of heat flux across the boundary. 

The core also transports heat by conduction or convection. For the core to convect, either the CMB heat flux, \fcmbp, must be superadiabatic or the core must be undergoing non-eutectic solidification. During inward core solidification, the density difference between the more dense, solidified pure iron and the less dense Fe-FeS liquid drives core convection, as the solidifed iron sinks to the centre of the core. When the core is convecting, the combined thermal and compositional buoyancy flux can be used to calculate a magnetic field strength \citep[see][for details]{sanderson_unlocking_2025}. Once the eutectic composition is reached, core solidification no longer generates a compositional buoyancy flux. Combining buoyancy flux contributions enables the thermal and compositional drivers of convection to be considered simultaneously.
The model ends once the whole core has solidified and dynamo generation is no longer possible. 

\subsubsection{Dynamo generation} \label{methods-Rem}
The vigour of convection in the liquid part of the core is characterised by the magnetic Reynolds number, $Re_m$, which must be supercritical for the core flow to be strong enough to generate a magnetic field. The analytical estimate of the critical value is 10, while from empirical simulations this value ranges from 40--100 \citep{stevenson_planetary_2003,aubert_modelling_2009}. We use a critical \Rem of 10 in line with previous work \citep{bryson_constraints_2019,dodds_thermal_2021}. However, the proportions of model runs with $10\leq\rm Re_m <40$, $40\leq\rm Re_m <100$, and $Re_m\geq100$ are also investigated. 

\subsection{Parameter variation}\label{methods-params}

Models were run for a range of mantle viscosities, planetesimal radii, initial core sulfur contents, \Xsp, and \feratio ratios in the accreted material (Table \ref{tab:variables}). 

\begin{figure}
    \centering
    \includegraphics[width=0.4\columnwidth]{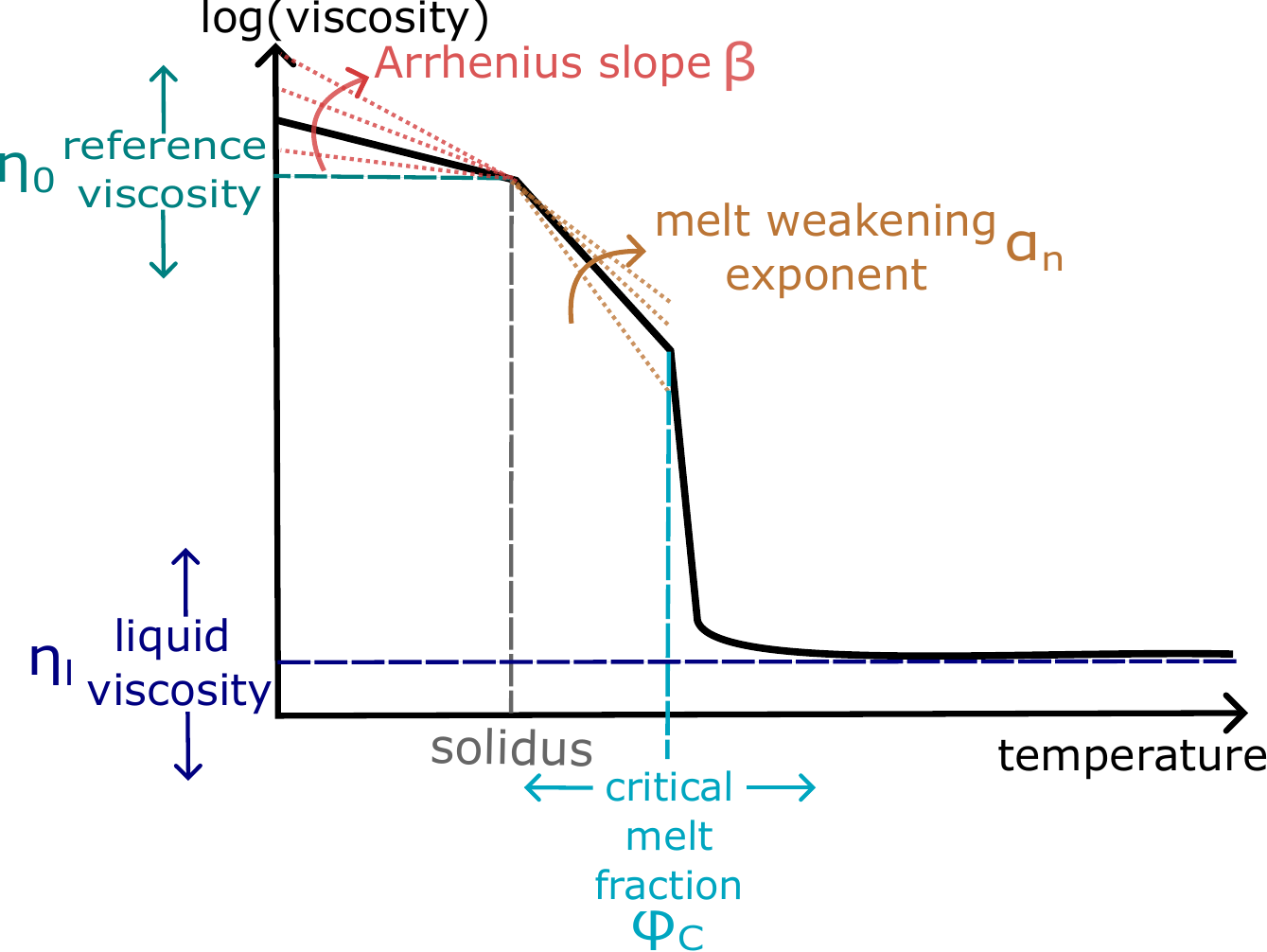}
    \caption{Illustration of parameters varied in the mantle viscosity model. Reference viscosity, \etarp=$10^{15}$--$10^{24}$\,Pas, is the viscosity at the mantle solidus. Liquid viscosity, $\eta_l$=1--100\,Pas, is the viscosity when the mantle is fully molten. Critical melt fraction, \rcmf=0.2--0.5, is the melt fraction for which all crystals become surrounded by melt, the material disaggregates and becomes rheologically weak. Arrhenius slope, $\beta$=0.01--0.035\,$\rm K^{-1}$, determines the rate of decrease in viscosity with increasing temperature. Melt weakening exponent, $\alpha_n$=25--40, determines the additional rate of decrease in viscosity with increasing temperature due to melting. }
    \label{fig:viscosity}
\end{figure}

The temperature dependence of mantle viscosity is defined piece-wise (Figure \ref{fig:viscosity}) by five control parameters. Below the solidus, the viscosity has an Arrhenius temperature dependence with slope $\beta$, which is proportional to the activation energy for viscous deformation. At the solidus, the mantle has a reference viscosity, \etarp. Above the solidus, melt weakens the material and the viscosity decreases with a steeper Arrhenius dependence, quantified by the melt weakening exponent, $\alpha_n$ \citep{hirth_rheology_2003}. At the critical melt fraction, \rcmfp, there is enough melt to completely surround any remaining solid phases and the material disaggregates with a rapid drop in viscosity. Beyond \rcmfp, the viscosity tends to a constant liquid viscosity, $\eta_l$, following the Krieger-Dougherty relation \citep{sturtz_birth_2022} as the melt fraction tends to one. 

Each parameter was varied independently while the others were held constant at the median value of their range, except \rcmf where the constant value was the experimental upper bound. The justifications for each parameter range are detailed in the following subsections.

\subsubsection{Reference viscosity, \etar} 
The reference viscosity, \etarp, ranges from $10^{15}$--$10^{24}$\,Pas. The lower bound is based on experimental measurements extrapolated to a range of mantle grain sizes \citep{scott_effect_2006}, while the upper bound is set by extrapolating reference viscosities used in models of Ganymede \citep{ruckriemen_top-down_2018} to planetesimal mid-mantle pressures (see Appendix \ref{extrap}). Due to the uncertainty in extrapolating from experimental to planetesimal conditions and the uncertainty in mantle grain sizes, the lower range of values from $10^{15}$--$10^{18}$\,Pas  are less likely to occur in planetesimals, but are included for completeness.

\subsubsection{Critical melt fraction, \rcmf} 
Experiments suggest the critical melt fraction, \rcmfp, is between 0.2--0.3 \citep{scott_effect_2006}. However thermal evolution models of planetesimals both for dynamo generation \citep{SterenborgCrowley2013,bryson_constraints_2019,dodds_thermal_2021} and differentiation \citep[e.g.][]{lichtenberg_magma_2019,monnereau_differentiation_2023} use a value of \rcmfp=0.5. Although it is unclear if \rcmf=0.5 is physically reasonable, we adopted the range \rcmf=0.2--0.5 to enable comparison with previous models and encapsulate experimental observations. 

\subsubsection{Arrhenius slope, $\beta$} 
The Arrhenius slope in the viscosity law, $\beta$, is typically defined as $\frac{E}{RT_{\text{ref}}^2}$, where $E$ is the activation energy, $R=8.31$\,J$\rm mol^{-1} K^{-1}$ is the gas constant, and $T_{\text{ref}}$ is a reference temperature - often taken to be the temperature at which the reference viscosity is measured. Experimentally measured activation energies for a range of silicate mantle compositions and strain rates range from 240--570 kJ mol$^{-1}$ \citep{karato_rheology_1993,hirth_rheology_2003}. For a reference temperature of the mantle solidus (1400\,K), this corresponds to $\beta$=0.015--0.035$K^{-1}$. \citet{dodds_thermal_2021} used $\beta = 0.01\,K^{-1}$, so to enable comparison with previous thermal evolution models, we explore $\beta =$ 0.01--0.035\,$K^{-1}$.

\subsubsection{Melt weakening exponent, $\alpha_n$}
The melt weakening exponent, $\alpha_n$, depends on whether deformation occurs by diffusion creep ($\alpha_n=$ 25--30) or dislocation creep ($\alpha_n=$ 30--45) \citep{hirth_rheology_2003}. To cover both deformation modes we consider values from 25--45.

\subsubsection{Liquid viscosity, $\eta_l$} 
The liquid viscosity, $\eta_l$, is determined by mantle composition and can range from 1--100\,Pas across the compositions of all volcanic rocks commonly encountered on Earth \citep{giordano_viscosity_2008}. Increasing silica content increases viscosity and increasing volatile content decreases viscosity \citep{lesher_chapter_2015}.  

\subsubsection{Initial core sulfur content, \Xs} 
The initial core sulfur content, \Xsp, ranges from the minimum Fe-FeS sulfur content for which the metal phase will be fully molten before differentiation to the eutectic composition (33\,wt\% for $T_{s,Fe}=1260$\,K for R=300\,km). The Fe-FeS liquidus temperature is pressure dependent so the minimum Fe-FeS changes with planetesimal radius.

\subsubsection{Primordial \feratio} 
Estimates on the primordial \feratio vary by almost three orders of magnitude from $10^{-9}$ \citep{fang_dating_2024} to $6\times10^{-7}$ \citep[e.g.][see Section S1 for more details]{cook_iron_2021}. However, previous thermal evolution and dynamo models neglected the presence of \fe in the core. Therefore, the model was run for five values of \feratio$=[0, 1\times10^{-9},1\times10^{-8}, 1\times10^{-7},6\times10^{-7}]$ to cover the full range of possible values and to compare with previous models. 

\subsubsection{Planetesimal radius, r}
Models were run for planetesimal radii between 100--500\,km. The lower limit is set by previous estimates for the minimum radius for compositional dynamo generation \citep{nimmo_energetics_2009}. The upper limit corresponds approximately to the radius of Ceres, the largest known asteroid today.
\begin{table*}
    \centering
    \begin{tabular}{|c|c|c|c|c|}\hline
        Symbol & Meaning & Range of values & Constant value & Reference \\\hline
        \etar & Reference viscosity & $10^{15}$--$10^{24}$\,Pas & $10^{19}$\,Pas & a, b \\
       \rcmf & Critical melt fraction & 0.2--0.5 & 0.3 & a, c\\
       $\beta$ & Arrhenius slope & 0.01--0.035\,$K^{-1}$ & 0.0225\,$K^{-1}$ & a, d, e \\
      $\alpha_n$ & Melt weakening exponent & 25--45 & 30 & f\\
      $\eta_l$ & Liquid viscosity & 1, 10, 100\,Pas & 10\,Pas & g \\\hline
       \Xs & Initial core sulfur content & 26.7--33\,wt\% & 29.85\,wt\% &  \\
       \feratio & \feratio in accreting material & 0, $10^{-9}$, $10^{-8}$, $10^{-7}$, $6\times10^{-7}$ & $10^{-8}$ & h, i, j, k \\
       r  & Planetesimal Radius & 100--500\,km & 300\,km & c\\\hline
    \end{tabular}
    \caption{Values for variable and constant parameters in simulations as described in Section \ref{methods-params}. The upper five parameters occur in the viscosity law. The range on \Xs is for a 300\,km radius planetesimal. For [500, 400, 200, 100]\,km radius planetesimals the minimum and median \Xs were [25.9, 26.3, 27.1, 27.1]\,wt\% and [29.45, 29.65, 30.05, 30.05]\,wt\% respectively. The constant values are median values in their range, except \rcmf where the constant value was the experimental upper bound. $\eta_l$ and \feratiop vary over several orders of magnitude, so the explored input values are stated explicitly and the constant value was the median from this list of values. (a) \citet{scott_effect_2006}, (b) \citet{ruckriemen_top-down_2018}, (c) \citet{bryson_constraints_2019}, (d) \citet{karato_rheology_1993}, (e) \citet{dodds_thermal_2021}, (f) \citet{hirth_rheology_2003}, (g) \citet{giordano_viscosity_2008}, (h) \citet{tang_abundance_2012}, (i) \citet{cook_iron_2021}, (j) \citet{kodolanyi_early_2022}, (k) \citet{kodolanyi_iron-60_2022}.}
    \label{tab:variables}
\end{table*}
\section{Results} \label{results}
\begin{figure*}
    \centering
    \includegraphics[width=1\textwidth]{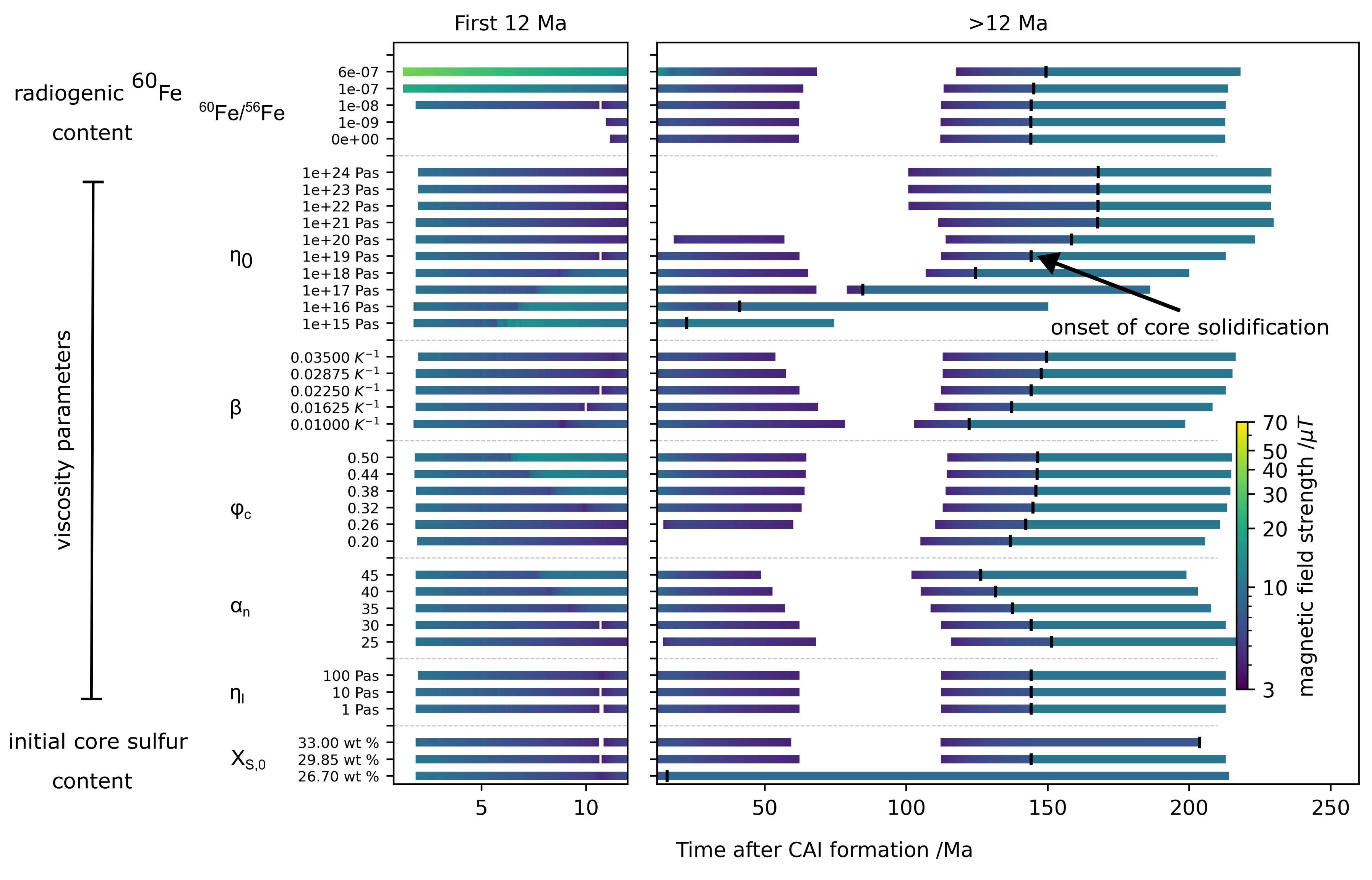}
    \caption{Dynamo timings for a 300\,km planetesimal for a range of initial \feratiop, mantle viscosities, and core compositions. Filled bars indicate periods when a dynamo is active and the colour of the bar indicates dipole magnetic field strength at the surface. The small, vertical, black line on each bar indicates the onset of core solidification in each model run. Magnetic field strengths between the onset of core solidification and the end of dynamo generation are averages for this time interval to remove oscillations in field strength due to the discretisation of the model \citep{sanderson_unlocking_2025}. The thin, vertical, white lines in some runs are $<0.5$\,Ma breaks in dynamo generation due to a rapid increase in viscosity at \rcmfp. These lines are artefacts of our viscosity and stagnant lid parametrisation and should not be interpreted. For each parameter, the model was run across the range of values in Table \ref{tab:variables}, while all other parameters are held at a constant median value (except \rcmf where the constant value was the experimental upper bound). The constant parameter values  were: $\eta_0=10^{19}$\,Pas, $\beta=0.0225\,K^{-1}$, $\alpha_n=30$, \rcmf= 0.3, $\eta_l=10$\,Pas, \feratio= $10^{-8}$, $X_{S,0}=29.85$\,wt\%. Magnetic field strengths as a function of time for each combination of parameters shown in this plot are provided in the Supplementary Materials.}
    \label{fig:dur}
\end{figure*}
\begin{figure*}
    \centering
    \includegraphics[width=1\textwidth]{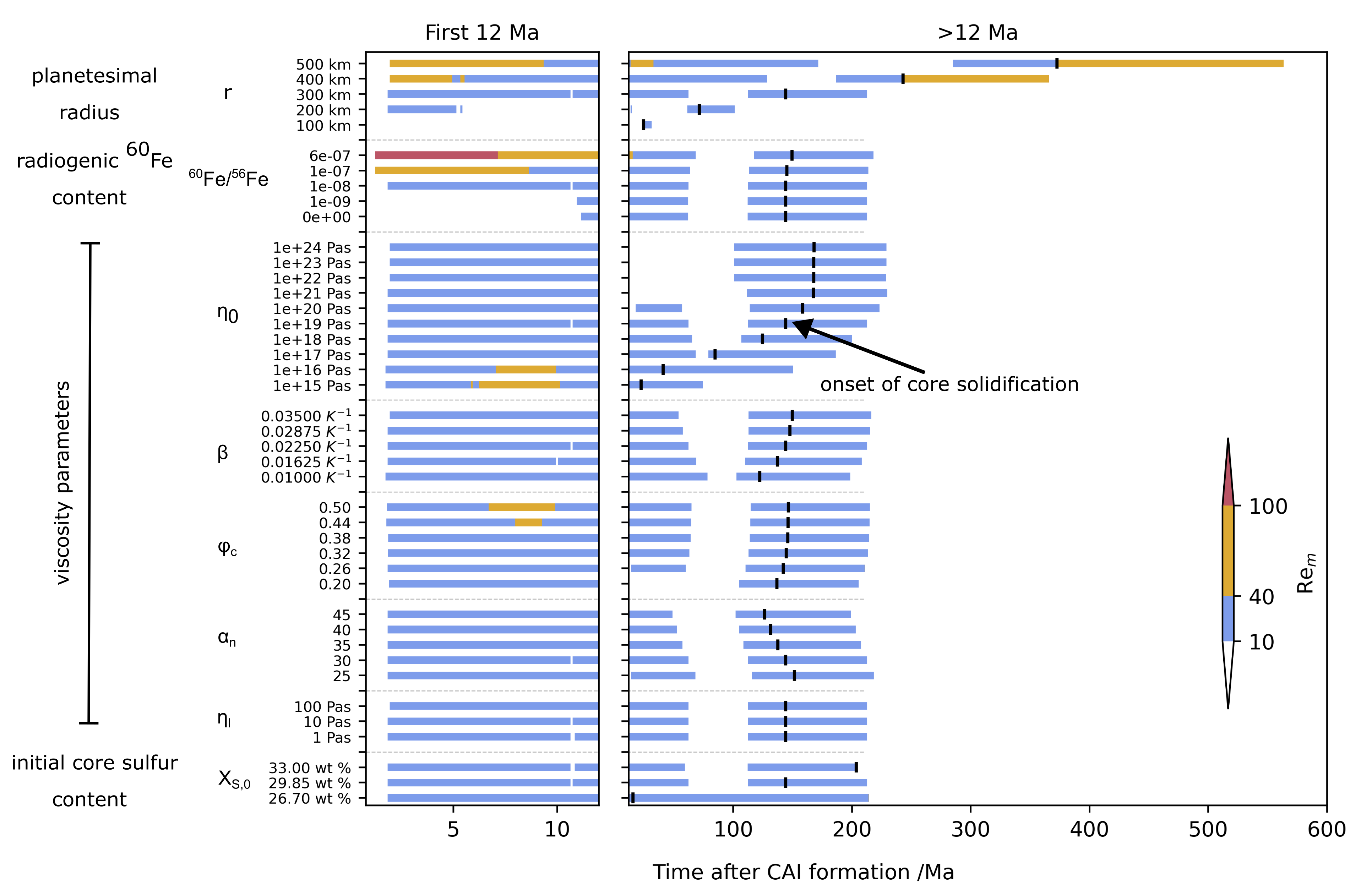}
    \caption{Magnetic Reynolds number ($Re_m$) as a function of time for a range of initial \feratiop, mantle viscosities, core compositions, and planetesimal radii. Filled bars indicate periods when the dynamo is on and the colour of the bars indicates the magnetic Reynolds number. Subcritical values ($<10$) are not shown. Supercritical values are categorised according to the range of possible criticality limits: 10-40, 40-100 and $>$100 \citep{christensen_scaling_2006,stevenson_planetary_2003}. The small, vertical, black line on each bar indicates the onset of core solidification in each model run. Magnetic Reynolds numbers between the onset of core solidification and the end of dynamo generation are averages for this time interval to remove oscillations in \Rem due to the discretisation of the model \citep{sanderson_unlocking_2025}. The thin, vertical, white lines in some runs are $<0.5$\,Ma breaks in dynamo generation due to a rapid increase in viscosity at \rcmfp. These lines are artefacts of our viscosity and stagnant lid parametrisation and should not be interpreted. For each parameter, the model was run across the range of values in Table \ref{tab:variables} and shown on Figure \ref{fig:dur}, while all other parameters were held at a constant median value (except \rcmf where the constant value was the experimental upper bound). The range of values for radius are shown explicitly. The constant parameters values  are: $r=300$\,km, $\eta_0=10^{19}$\,Pas, $\beta=0.0225\,K^{-1}$, $\alpha_n=30$, \rcmf= 0.3, $\eta_l=10$\,Pas, \feratio= $10^{-8}$, $X_{S,0}=29.85$\,wt\%.}
    \label{fig:Rem}
\end{figure*}

\subsection{Constant radius}

Firstly, we ran models with a constant radius of 300\,km to identify the impact of other parameters on thermal evolution and dynamo generation. The lowest mantle reference viscosity, \etarp, and lowest initial core sulfur content, \Xsp, result in a single epoch of dynamo generation, and a single value in each of \etarp, \rcmfp, and $\alpha_n$ result in three epochs of dynamo generation (see Supplementary Materials). Otherwise, all other parameter combinations produce two periods of dynamo generation (Figure \ref{fig:dur}). Across these two dynamo generation periods, we observe two distinctive trends in magnetic field strength: (i) the first dynamo generation period can have two peaks in magnetic field strength ii) in the second dynamo generation period, magnetic field strength increases with time towards a peak value during core solidification. The first peak in the first epoch of dynamo generation corresponds to the onset of dynamo generation and the second peak occurs when the base of the mantle first cools below the critical melt fraction \citep{sanderson_unlocking_2025}. The peak in field strength at the beginning of the first epoch of dynamo generation is always the strongest because radiogenic heating by \fe increases CMB heat flux (\fcmbp). For all runs with multiple field generation periods, the final dynamo generation period begins before the onset of core solidification. This is due to the delayed steepening of the mantle conductive gradient at the CMB after the cessation of mantle convection, which increases \fcmb \citep{sanderson_refined_2024}. Reference viscosity has the strongest effect on the duration of the gap in dynamo generation and \feratio has the strongest influence on the resulting magnetic field strength. Despite the range in dynamo durations and strengths, most \Rem values fall between 10--40 (Figure \ref{fig:Rem}).

\subsubsection{Effect of $^{60}Fe$}
Increasing \feratio brings forward the onset of the dynamo to 1--2\,Ma after CAI formation and increases peak field strength and \Rem in the first epoch of dynamo generation. The onset of the dynamo is earlier for increased \fe because increased heating either erodes core thermal stratification more rapidly or prevents its formation altogether, which brings forward the time that \fcmb becomes superadiabatic. Radiogenic heating by \fe also increases the temperature of the core and the temperature difference relative to the mantle, which increases \fcmb and the strength of the dynamo. Model runs with \feratio$\geq 10^{-7}$ had peak field strengths in the first period of dynamo generation 2--3 times higher than runs with \feratio$=10^{-8}$. This extra heat is removed from the core by an elevated \fcmb during the first period of dynamo generation. As such, by the second epoch of dynamo generation the peak field strengths, \Rem and the onset of core solidification are similar across all values of \feratiop. There are similar onset times and field strengths for the first dynamo generation period for \feratio=0 and \feratio$=10^{-9}$. This similarity suggests that when \feratio$\leq10^{-9}$ the effect of \fe can be neglected. The onset times vary little across variations in other parameters indicating \feratio is the dominant control on onset time for the first period of dynamo generation.

\subsubsection{Effect of viscosity}
The reference viscosity, \etarp, has the strongest effect on the duration of the first period of dynamo generation and the following gap. Above $10^{21}$ Pas, the first period of dynamo generation ends after 13--14\,Ma and is followed by a gap of >80\,Ma in dynamo generation. This results from the rapid thickening of the surface stagnant lid shutting off convection earlier for the highest reference viscosities. In contrast, below $10^{16}$\,Pas, there is only one epoch of dynamo generation, because the mantle viscosity is sufficiently low that mantle convection does not cease before core solidification. The second peak in magnetic field strength in the first epoch of dynamo generation widens and increases in strength with decreasing reference viscosity, because the thinner boundary layer at the CMB raises \fcmbp. An elevated \fcmb increases the rate at which the core cools, which brings forward the onset of core solidification and the end of dynamo generation. 

The viscosity parameters have very little effect on the supercriticality of \Remp. Only the lowest \etar and highest values of \rcmf increase \Rem in the first 12\,Ma (Figure \ref{fig:Rem}). However, the highest value \rcmf=0.5 has been adopted in previous thermal evolution models, despite not being supported by experimental data for basaltic melts \citep{scott_effect_2006}. The elevated \Rem for this value may have resulted in overestimates of field strength and dynamo duration in previous work. The lowest values of \etar ($<10^{18}$\,Pas are less applicable, because of the uncertainty in extrapolating from laboratory to planetesimal grain sizes (see Appendix \ref{extrap}). As such, the very rapid core cooling and elevated \Rem observed in models with \etar$\leq10^{16}$\,Pas is unlikely to occur in planetesimals.

\subsubsection{Effect of \Xs}
Intial core sulfur content, \Xsp, has a strong effect on the timing of core solidification, because increasing \Xs lowers the liquidus temperature of the core and more cooling is required to reach this temperature. The lowest \Xs removes the gap in dynamo generation, because the early onset of core solidification provides additional compositional buoyancy flux to drive the dynamo even when mantle convection ceases.

\subsection{Variable radius}
\begin{figure*}
    \centering
    \includegraphics[width=1\textwidth]{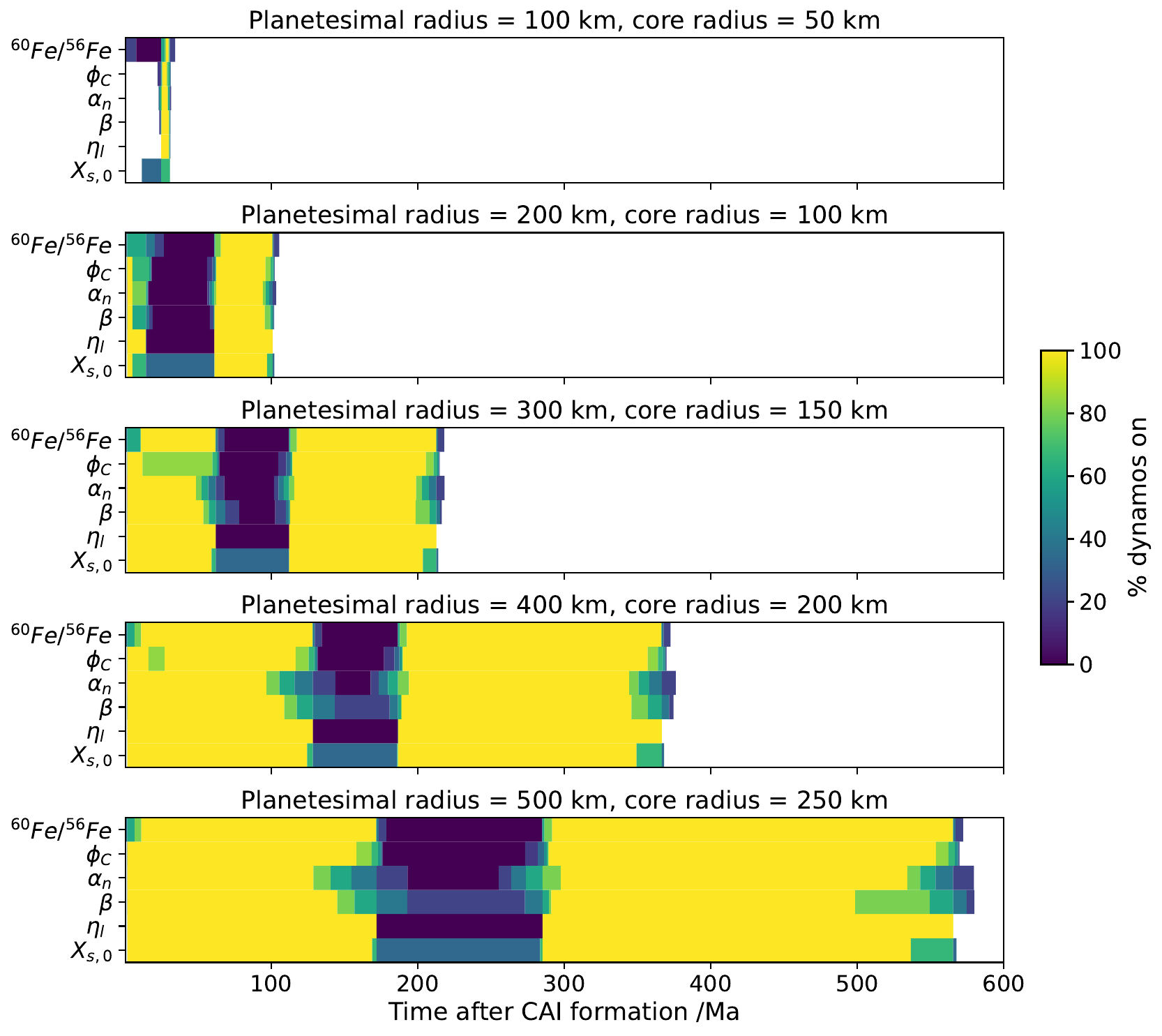}
    \caption{Percentage of active dynamos across the range of possible values for initial \feratiop, mantle viscosities, and core compositions for planetesimals ranging from 100-500\,km in radius. In an individual panel, model runs are grouped in horizontal blocks by parameter. Yellow indicates that all dynamos are on at a given time irrespective of the value of a parameter, dark blue indicates no dynamos are on at a given time irrespective of the value of a parameter. Light blue to green regions indicate a dynamo is on for a subset of values of that parameter. White space indicates times after the last period of dynamo generation and the end of the simulation. The constant parameters values  are: $\eta_0=10^{19}$\,Pas, $\beta=0.0225\,K^{-1}$, $\alpha_n=30$, \rcmf=0.3, $\eta_l=10$\,Pas, \feratio=$10^{-8}$ and the range for each parameter is given in Table \ref{tab:variables}. $\eta_0$ causes a wide variation in dynamo timings, so is shown separately in Figure \ref{fig:eta0}.}
    \label{fig:heatmap}
\end{figure*}
\begin{figure*}
    \centering
    \includegraphics[width=1\textwidth]{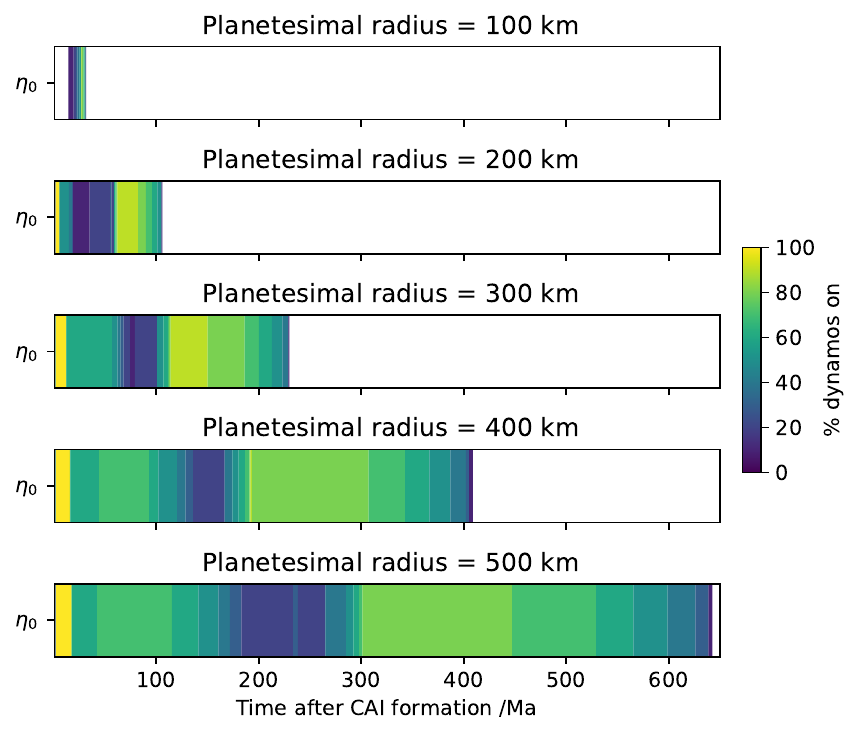}
    \caption{Percentage of active dynamos across the range of possible values for reference viscosity, \etarp, for planetesimals ranging from 100--500\,km in radius. Core radius is half planetesimal radius. Yellow indicates that all dynamos are on at a given time irrespective of the value of \etarp, dark blue indicates no dynamos are on at a given time irrespective of the value of \etarp. Light blue to green regions indicate a dynamo is on for a subset of values of \etarp. White space indicates times after the last period of dynamo generation and the end of the simulation. The vertical axis is equivalent to one horizontal block in Figure \ref{fig:heatmap} and has no physical meaning. The constant parameter values are: $\beta=0.0225\,K^{-1}$, $\alpha_n=30$, \rcmf= 0.3, $\eta_l=10$\,Pas, \feratio= $10^{-8}$, $X_{S,0}=$[29.45, 29.65, 29.85, 30.05, 30.05]\,wt\% for [500, 400, 300, 200, 100]\,km radius planetesimals. The range for \etar is given in Table \ref{tab:variables}.}
    \label{fig:eta0}
\end{figure*}
To identify the impact of planetesimal radius on thermal evolution and dynamo generation we ran models with varying radii and other variables held constant (Table \ref{tab:variables}). Core radius was assumed to be half the planetesimal radius. Planetesimal radius has the strongest effect on dynamo timing, duration, and width of the gap in dynamo generation (Figure \ref{fig:heatmap}). Larger planetesimals cool more slowly so mantle convection ceases later, which prolongs the first epoch of dynamo generation. Larger core size increases the lengthscale of convection; this raises the \Rem value for a given \fcmbp, which further prolongs dynamo generation. Larger core size also increases magnetic field strengths (Figure S1--S4). The increase in magnetic field strength is most pronounced at the onset of the first epoch of dynamo generation for when the highest values of \feratio are considered. Bodies with radius of 100\,km only generate magnetic fields after the onset of core solidification (Figure \ref{fig:Rem}) except for the highest values of \feratiop. Increasing planetesimal radius from 100\,km to 500\,km increases the end time of dynamo generation from $\sim$30 to $\sim$600\,Ma. The gap in dynamo generation moves later ($\sim$20--200\,Ma) and increases in duration ($\sim$20--270\,Ma) for increasing planetesimal radius. For each body size, the timings of the dynamo generation periods are consistent across variations in the other parameters (Figure \ref{fig:heatmap}). An exception is \etar (Figure \ref{fig:eta0}), where values <$10^{16}$\,Pas do not produce a gap in dynamo generation. However, there is still a trend of increasing dynamo duration with radius for \etarp.

\section{Discussion} \label{discussion}
\begin{figure*}
    \centering
    \includegraphics[width=1\textwidth]{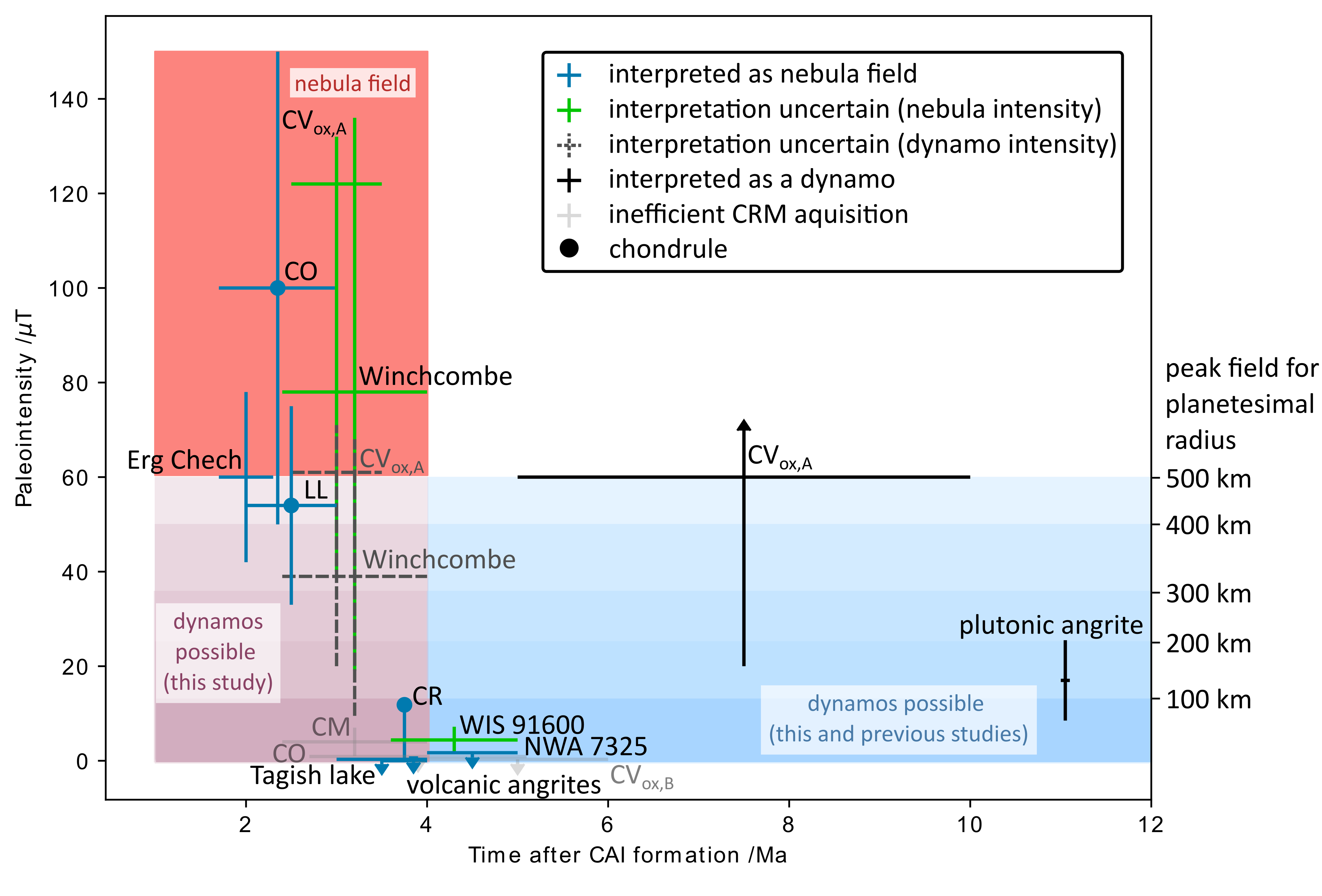}
    \caption{A summary of meteorite paleointensities for the first 12\,Ma after CAI formation. Meteorites for which the cause of magnetisation is interpreted as a nebula field (or lack of) are plotted in blue. Meteorites for which the cause of magnetisation has been interpreted as a dynamo field are plotted in black. Meteorites which could have a nebula or dynamo origin are plotted in green for their possible nebula field paleointensities. The dashed grey lines indicate the paleointensities of Winchcombe \citep{bryson_unified_2023} and Allende \citep{fu_fine-scale_2021} if the factor of two correction for the nebula field is not applied. Faded grey points correspond to null or low paleointensities, which may be due to formation of magnetite by pseudomorphic replacement which cannot record a chemical remnant magnetisation (CRM) \citep{bryson_unified_2023}. The blue circles indicate pre-accretionary remanences from chondrules.  The IVA irons have not been included, because they were generated by a solidifying, unmantled core \citep{bryson_paleomagnetic_2017}. The R chondrites have too large an age uncertainty and the CV chondrules are too altered to provide useful constraints, so neither are shown \citep{cournede_paleomagnetism_2020,shah_long-lived_2017}. Intensity bars with arrows indicate values with a upper/lower limit (the cap) and for the $\rm CV_{ox,A}$ the best guess is the intersection with the time error bar. Our new model brings forward the onset of dynamo generation (pink box) relative to previous models (blue box). The graduated shading in the blue and pink regions indicates the maximum paleointensity for a given planetesimal radius (shown by ticks on the right hand vertical axis). The orange box indicates the duration of the nebula field, which may be 1\,Ma longer than shown here if the angrite age is recalculated based on \citet{piralla_unified_2023}. This would further increase the overlap between dynamo and nebula fields.  Data and references for this table are given in the Supplementary Materials.}
    \label{fig:new-paleomag}
\end{figure*}
\subsection{Reevaluating the paleomagnetic record}
The purpose of our model is to determine the origins (nebula field, thermal or compositional dynamo) of magnetisations in the meteorite paleomagnetic record. Overall, our model results suggest two key interpretations of the paleomagnetic record should be revisited. First, the earliest recorded magnetisations ($<$5\,Ma after CAI formation), previously attributed to the nebula field, could have been dynamo generated. This is due to inclusion of \feratio $\geq 10^{-8}$ in our model bringing forward the onset of thermal dynamos to 1--2\,Ma after CAI formation , which allows for thermal dynamo generation prior to the dissipation of the nebula field (Figure \ref{fig:new-paleomag}). Second, late magnetic remanences, previously ascribed to the second epoch of dynamo generation during core solidification, may not have been in the second epoch and do not require core solidification, but could instead be produced by a thermal dynamo. Our model predicts an elongated first epoch of dynamo generation lasting until 20--200\,Ma after CAI formation (depending on radius) rather than until 9--34\,Ma after CAI formation as suggested by previous models \citep{dodds_thermal_2021}. Additionally, in our model, the second epoch of dynamo generation begins before the onset of core solidification. The meteorite paleomagnetic record in light of our new model results is reassessed in detail below.

\subsubsection{Paleomagnetic records of the nebula field}
 
Unidirectional magnetisation within chondrules, but non-unidirectional magnetisation between chondrules in a meteorite is only possible if the chondrules acquired their remanence prior to accretion onto the meteorite parent body. The nebula field is the only source of a pre-accretionary magnetisation, therefore non-unidirectional chondrule magnetisations provide conclusive evidence that a paleointensity can be attributed to the nebula field. Non-unidirectional measurements of chondrules from CO \citep{borlina_paleomagnetic_2021}, LL3 \citep{fu_solar_2014} and CR chondrites \citep{fu_weak_2020} are interpreted as evidence for the nebula field $\sim$1.7--4\,Ma after CAI formation with paleointensities up to $\sim\!150\,\mu$T (Figure \ref{fig:new-paleomag}). Bulk meteorite magnetisations from parent bodies that were too small to generate a dynamo can also be attributed to the nebula field \citep[e.g. Erg Chech 002,][]{neumann_fitting_2023,maurel_4565-my-old_2024}.


Null remanences could indicate the dissipation of the nebula field, but there are two other possible interpretations, which must be excluded first. First, a meteorite parent body may have formed in a distal part of the Solar System where the field was weak \citep[e.g. Tagish Lake,][]{bryson_evidence_2020}. Second, the magnetic minerals may not have formed in a way that allowed them to record a remanence. Many carbonaceous chondrites record magnetisation in magnetite, which 
formed during aqueous alteration. If the magnetite formed by pseudomorphic replacement of FeNi metla rather than precipitation, it may not have recorded a new remanence during this process \citep{bryson_unified_2023}.
This could explain the null remanence in CV chondrite Kaba \citep{gattacceca_new_2016} and the upper limit on paleointensity of $0.9\,\mu$T from the matrix of CO chondrites \citep{borlina_lifetime_2022}. For the volcanic angrites both these interpretations can be excluded. They formed in the inner Solar System, erupted $\sim$3.8--3.9\,Ma after CAI formation\footnote{These ages were originally reported as absolute Pb-Pb ages and may be shifted to 4.8--4.9\,Ma after CAI formation based on reevaluation of the anchor for Pb-Pb \citep{piralla_unified_2023}.} and cooled quickly forming magnetite that should have recorded any magnetic fields present \citep{wang_lifetime_2017}. Therefore, a null remanence in the angrites suggests the nebula field in the inner Solar System dissipated 3.8--3.9\,Ma after CAI formation \citep{wang_lifetime_2017}. The recovered paleointensity of $<0.6\,\mu$T from ungrouped achondrite NWA 7325 at $\sim$4--5\,Ma after CAI formation \citep{weiss_nonmagnetic_2017} further supports the nebula field in the inner Solar System having dissipated by this time. 

\subsubsection{Ambiguous paleomagnetic records: nebula or dynamo fields?}

Interpretation of bulk chondrite paleomagnetic remanences (i.e. not individual chondrule measurements) from the first 5\,Ma after CAI formation is complicated because these meteorites could sample the outer layer of a partially differentiated planetesimal \citep{carporzen_magnetic_2011,elkins-tanton_chondrites_2011,weiss_differentiated_2013}. In this case, a chondritic meteorite could record a paleomagnetic remanence from a dynamo generated in the core of their partially differentiated parent body rather than the nebula field. The feasibility of this scenario rests on the timing of planetesimal dynamos, evidence for partially differentiated meteorite parent bodies, and the value of primordial \feratiop. Our model demonstrates that a parent body, with \feratio$\geq10^{-8}$, would have an early onset of a thermal dynamo before the dissipation of the nebula field. Thermally metamorphosed chondritic meteorites could have been heated purely radiogenically on a parent body that partially differentiated. The interpretation of bulk remanences in the CM and CV chondrites and WIS 91600 (green points in Figure \ref{fig:new-paleomag}) is less clear than previously, in light of the possibility they record an early dynamo. The ages of the R chondrites are too poorly constrained \citep[0--20\,Ma after CAI formation,][]{cournede_paleomagnetism_2020} for the origin of their remanence to be discussed.

The CM chondrites and WIS 91600 recorded a magnetic field 2--4\,Ma after CAI formation \citep{cournede_early_2015,bryson_unified_2023} and likely 3.6--5\,Ma after CAI formation \citep{bryson_constraints_2020}, respectively. The timing of these remanences is consistent with either a nebula field or dynamo origin. For example, a 500\,km radius planetesimal with \feratio=1--6$\times10^{-7}$ and other constant parameters from Table \ref{tab:variables} produces intensities that match those recovered from Winchcombe  (Figure S10). However, lack of evidence for prolonged thermal metamorphism in the CM chondrites and WIS 91600 \citep{suttle_aqueous_2021,bryson_constraints_2020} argues against partially differentiated parent bodies for these meteorites; unless they sample the topmost, coolest layer of such a body. Further numerical modelling of partially differentiated bodies combining gradual accretion and \fe are required to determine whether a dynamo in a partially differentiated body or the nebula field is the source of these remanences. 

The CV chondrites are thermally metamorphosed \citep{carporzen_magnetic_2011,elkins-tanton_chondrites_2011,gattacceca_new_2016} and have a complex paleomagnetic record. CV chondrules have been modified by impacts, heating and aqueous alteration and have a non-unidirectional magnetisation within an individual chondrule, so their remanence cannot be attributed to the nebula field \citep{fu_no_2014,shah_long-lived_2017}. Allende records a magnetisation in the iron sulfides in the matrix from 2.5--3.5\,Ma after CAI formation \citep{fu_fine-scale_2021}, and has a bulk remanance dated to 5--40\,Ma after CAI formation \citep{carporzen_magnetic_2011,elkins-tanton_chondrites_2011}. The CV chondrite Kaba has a null remanence from aqueous alteration at 4--6\,Ma after CAI formation, but a $\sim$3\,$\mu$T remanence from prolonged metamorphism tens of Ma after CAI formation \citep{gattacceca_new_2016}. The later ($>5$\,Ma after CAI formation) CV chondrite remanences, combined with evidence for thermal metamorphism, suggest the CV chondrite parent body was partially differentiated and generated a dynamo \citep{elkins-tanton_chondrites_2011,gattacceca_new_2016}. Previously, Kaba's null remanance, recorded during aqueous alteration, was used as evidence for the dissipation of the nebula field before the onset of the dynamo on the CV chondrite parent body. This supported a nebula origin for the earliest Allende remanence \citep[2.5--3.5\,Ma after CAI formation][]{fu_fine-scale_2021}. However, Kaba's magnetite may have been unable to acquire a remanance during formation \citep{bryson_unified_2023}, so this null remanance may not indicate an absence of a magnetic field on the CV parent body at 4--6\,Ma after CAI formation. Altogether, the altered interpretation of Kaba's null remanence, thermal metamorphism of the CV chondrite parent body and the early dynamo onset predicted by our models suggests the earliest remanence on Allende could be due to the early onset of a thermal dynamo. 

\paragraph{Intensity of remanence}
Next, we assess whether paleointensity can be used to determine if a remanence has a dynamo or nebula field origin. As shown in Figure \ref{fig:new-paleomag}, the maximum nebula intensities measured are stronger than the maximum intensities predicted by our dynamo model. However, the quoted nebula intensities recorded by Winchcombe \citep[CM chondrite,][]{bryson_unified_2023} and Allende \citep[CV chondrite,][]{fu_fine-scale_2021} are a factor of two larger than the measured paleointensities to account for rotation of the parent body during prolonged remanence acquisition. If these paleointensities were produced by dynamo fields this correction would not be required, because the magnetic field is co-rotating with the planetesimal. Removing this correction decreases their paleointensities so they are within the range of field strengths predicted by our dynamo models (dashed lines on Figure \ref{fig:new-paleomag}). Our upper limits on dynamo field strengths ($>30$\,$\mu$T) are for the largest planetesimals with highest values of \fe at early times ($<3$\,Ma after CAI formation). Most parameter combinations predict weaker dynamo strengths $<3$\,Ma after CAI formation (10--30\,$\mu$T, Figure \ref{fig:dur} and S1--S4). A combination of approaches would be required to determine if measured paleointensities, even without the factor of two amplification, are too strong to be dynamo generated. The uncertainty on the paleointensity measurements could be reduced, and parent body size and \feratio could be independently constrained, e.g. by thermal models and isotopic measurements. Regardless of improved measurements, the difficulty in converting from field strengths acquired by different mechanisms in the laboratory compared to the parent body \citep[factors of 2--5,][]{weiss_lunar_2014,maurel_estimating_2023} and uncertainties in the ratio of dipole field at the CMB to the internal dynamo field from dynamo scaling laws \citep[0.06--0.1 for Earth,][]{davies_dynamo_2022} may mean that the uncertainties of measured and predicted paleointensities overlap. Overall, current recovered paleointensities and modelled field intensities cannot be used to distinguish between a dynamo or nebula field.

The possibility of the earlier onset of a thermal dynamo due to \fe reduces the number of meteorite paleomagnetic studies that provide compelling evidence for the nebula field in the outer Solar System, because the bulk remanences in CM and CV chondrites may not record nebula fields. This should be considered when using this data to discuss differences in field strength and protoplanetary disk dispersal time between the inner and outer Solar System \citep{weiss_history_2021,borlina_lifetime_2022}. Paleomagnetic measurements of individual, pristine chondrules are vital to understand the nebula field, because their non-unidirectionality can indicate conclusively if a remanence is preaccretionary. 

\subsubsection{Paleomagnetic records of planetesimal core crystallisation}
Our results argue that late remanences ($>65$\,Ma after CAI formation based on the current meteorite record) do not require the presence of a solidifying core. The first epoch of dynamo generation can last until 20--200\,Ma after CAI formation (depending on radius) rather than until 9--34\,Ma after CAI formation suggested by previous models \citep{dodds_thermal_2021}. Additionally, our improved core solidification model reveals the second epoch of dynamo generation can begin prior to the onset of core solidification. As a result, the source of buoyancy to drive a dynamo cannot be assumed purely based on timing of a magnetic remanence. For extremes of some parameter values (low \etar and \Xsp), onset of solidification is early and most of dynamo generation is powered by a combination of thermal and compositional buoyancy. Previous studies of the Main Group pallasites \citep{bryson_long-lived_2015,nichols_time-resolved_2021} and IIE irons \citep{maurel_meteorite_2020} have used the onset of a late-stage dynamo as an indicator for the start of core solidification. This assumption has been used to constrain planetesimal radius, core size, thermal history, and core composition. Our model highlights that the beginning of a second epoch of dynamo is unlikely to be coincident with the beginning of core solidification for any initial core sulfur content, and therefore these studies should be re-evaluated.

\subsubsection{Paleomagnetic constraints on primordial $^{60}$Fe/$^{56}$Fe}\label{dis-angrite}
The null paleointensities in the volcanic angrites and NWA 7325 at 3.8--5\,Ma after CAI formation  \citep{wang_lifetime_2017,weiss_nonmagnetic_2017} contradicts the early onset of a dynamo in bodies with \feratio $\geq10^{-8}$ predicted by this model. Magnetic remanences in the plutonic angrite, Angra dos Reis at 11\,Ma after CAI formation \citep{wang_lifetime_2017}, indicate the angrite parent body was differentiated and did generate a dynamo later in its history.  Explanations, such as parent body size, gradual accretion, or \fe fractionation are unlikely to cause this delay in dynamo onset (see Supplementary Materials). Another explanation could be \feratio was $\leq10^{-9}$ in the formation region of the angrites such that the time taken to erode core thermal stratification led to a delay in dynamo generation \citep{bryson_constraints_2019,dodds_thermal_2021}. \citet{tang_abundance_2012} measured values of \feratio$=0.9\pm0.5\times10^{-8}$ in the angrites. Additionally, values of \feratio from silicates in non-carbonaceous chondrules are within error of zero \citep[$0.8\pm1.0\times10^{-7}$,][]{kodolanyi_early_2022}. For a 300\,km body accreted at 0.8\,Ma after CAI formation, \feratio$<1.75\times10^{-9}$ results in a dynamo onset time consistent with the angrite data (see Figure S5). This limit lies within the lower bounds of these non-carbonaceous \feratio values. Current measurements from the carbonaceous IID and IVB irons \citep[$6\pm2\times10^{-7}$,][]{cook_iron_2021} and carbonaceous chondrule silicates \citep[$4\pm2\times10^{-7}$][]{kodolanyi_early_2022} have large uncertainties (see Section S1), but neither are within error of \feratio $<10^{-8}$. This suggests there could be heterogeneity in \feratio between the carbonaceous and non-carbonaceous reservoirs \citep{cook_iron_2021,kodolanyi_early_2022}. Measurement of a paleomagnetic remanence in a carbonaceous achondrite between the age of the volcanic and plutonic angrites could inform this discussion. If such a carbonaceous achondrite has a null paleomagnetic remanence, this could provide evidence for \feratio$<10^{-8}$ throughout the Solar System and constrain the timing of the dissipation of the nebula field in the outer Solar System. If this carbonaceous achondrite is found to have acquired a remanence at a similar time to null remanence acquisition in the volcanic angrites (i.e., when the nebula field has dissipated), this could provide evidence for early dynamo onset and heterogeneity in \feratio in the early Solar System. 
 
\subsection{Viscosity parameters in future planetesimal thermal evolution models}
Mantle critical melt fraction, \rcmfp, is a crucial parameter in thermal evolution models. It determines planetesimal peak temperature and melt fraction (see Appendix \ref{rcmf}), and defines the threshold for magma ocean formation and convection in some models \citep[e.g.][]{sturtz_structure_2022,neumann_fitting_2023}. A range of values of \rcmfp, including lower values in line with experimental data \citep{scott_effect_2006} should be used in future models. For models of specific parent bodies, this will affect the extent of the magma ocean and the inferred initial location of a meteorite \citep[e.g. Erg Chech 002,][]{sturtz_structure_2022}. In two-phase flow models for planetesimal differentiation \citep[e.g.][]{lichtenberg_magma_2019,monnereau_differentiation_2023}, reducing \rcmf and planetesimal peak temperature may alter the Fe-FeS compositions that can be molten, which will affect initial core composition. Additionally, reducing the melt threshold for magma ocean onset in differentiation models may decrease the time available for processes such as differentiation via percolation.

The value of mantle reference viscosity, \etarp, has a strong effect planetesimal cooling rate. Future models should test multiple values in the extrapolated range of \etar when attempting to constrain a parent body thermal history from meteorite isotope data \citep[e.g.][]{bryson_paleomagnetic_2019,neumann_fitting_2023}. Future models could attempt to constrain values for planetesimal \etar (and other viscosity parameters) based on parent body composition, water content, and observed gaps in the paleomagnetic record. 

\subsection{Validity}
This work only explored variations within one parameter at a time, while all other parameters were held at their median values (except \rcmf where the constant value was the experimental upper bound). Using endmember values for multiple parameters simultaneously may result in an even wider range of dynamo generation times. Varying a single parameter enables the effect of each parameter to be understood and the dominant factors in dynamo generation timing to be isolated. As demonstrated in Figure \ref{fig:heatmap}, even across variations in parameters other than radius, there are large portions of time where the planetesimal dynamo is consistently active. Therefore, even if multiple parameters are varied, the trends in timing with radius predicted here are likely to hold true. The choice of critical \Remp, light elements in the core, \Xsp, core radius fraction and assumptions about accretion and differentiation are justified for the purposes of this work and do not have a large impact on the results (for further discussion see the Supplementary Materials). 

\section{Conclusions} \label{conclusion}
Using the refined planetesimal thermal evolution model of \citet{sanderson_unlocking_2025}, we explored the effects of \fep, mantle viscosity, initial core sulfur content, and planetesimal radius on planetesimal dynamo generation. Our results suggest there is a wider range in dynamo timings and durations than previously predicted. The meteorite paleomagnetic record therefore cannot be split into three distinct epochs of magnetic field generation (nebula field, thermal dynamos and compositional dynamos) and the implications of some paleomagnetic remanences should be re-evaluated.

\begin{itemize}
    \item Inclusion of \feratio $\geq10^{-8}$ in the core can bring forward the onset of thermal dynamos to 1--2\,Ma after CAI formation, extending the time at which the first planetesimal dynamos were possible. Therefore, some meteorite remanences recorded prior to 5\,Ma after CAI formation could have a thermal dynamo origin rather than a nebula field origin. As a result, individual, pristine chondrules are crucial to understand the nebula field, because their non-unidirectionality can indicate conclusively that a remanence is preaccretionary.
     \item Null paleomagnetic remanences in the angrites contradict the early onset of thermal dynamos caused by \fep. This could provide evidence for heterogeneous \feratio in the early Solar System. Paleomagnetic measurements of an old ($<$10\,Ma after CAI formation) carbonaceous achondrite could help inform this.
     \item Increasing planetesimal radius (from 100\,km to 500\,km) greatly increases the start time of the gap in planetesimal dynamo generation (20 to 200\,Ma after CAI formation) and final end of dynamo duration (40 to 640\,Ma after CAI formation).    
    \item The paleomagnetic record does not reflect early thermal dynamos and later compositional dynamos during core solidification, as previously interpreted. Variation in initial core sulfur content and mantle reference viscosity can produce early thermo-compositional dynamos. For most parameter combinations, there is a second thermal dynamo period prior to the onset of core solidification.
    \item Late magnetic remanences ($>$65\,Ma post CAI formation based on the current record) do not require a solidifying core. As a result, the planetesimal radii, core sizes, thermal histories, and core compositions of the Main Group pallasites and IIE irons should be re-evaluated.
    \item A range of planetesinal radii, mantle reference viscosities, and initial \feratio ratios should be explored when attempting to constrain parent body properties based on a meteorite paleomagnetic remanence. 
\end{itemize}

\printcredits 

\section*{Declaration of competing interest}
The authors declare that they have no known competing financial interests or personal relationships that could have appeared to influence the work reported in this paper.

\section*{Data availability}
The dynamo and thermal evolution model and the parameter files required to recreate the results in this paper are publicly available the \citet{sanderson_refined_2024} Github repository. 

\section*{Acknowledgements}
The authors would like to thank Kathryn Dodds and Chris Davies for their helpful discussions in preparing this work. HRS acknowledges funding on a NERC studentship NE/S007474/1 and an Exonian Graduate Scholarship from Exeter College, University of Oxford. HRS thanks the Royal Astronomical Society for a student travel grant that enabled this work to be presented at the American Geophysical Union Fall Meeting 2023. JFJB acknowledges funding from the UKRI Research Frontier Guarantee program EP/Y014375/1. For the purpose of open access, the authors have applied a Creative Commons Attribution (CC BY) licence to any Author Accepted Manuscript version arising. 

\appendix
\section{Extrapolation of reference viscosities}\label{extrap}
Viscosity is temperature, pressure and grain size dependent. Planetesimal mid-mantle pressures lie above that of lab experiments and below that of planets, therefore extrapolation is required when estimating planetesimal reference viscosities. Laboratory experiments are also on shorter timescales and smaller grain sizes compared to planetary conditions. Therefore, geophysical observations, such as postglacial uplift \citep{karato_rheology_1993} are often used to constrain mantle viscosity on Earth.  For this reason, the upper range of \etar in this paper was extrapolated from planetary values using an Arrhenius law \begin{equation}
    \eta = \eta_0exp\left(\frac{E+pV}{RT}-\frac{E+p_{\text{ref}}V}{RT_{\text{ref}}}\right)
\label{eq:arr}\end{equation} and the Frank-Kamenetskii approximation \citep{noack_first-_2013}\begin{equation}
  \eta = \eta_0exp\left(\frac{E+p_{\text{ref}}V}{RT_{\text{ref}}^2}(T_{\text{ref}}-T)+\frac{V}{RT_{\text{ref}}}(p-p_{\text{ref}})\right).
\label{eq:fk}\end{equation} 
In these equations, $E$ and $V$ are the activation energy and volume, $p_{\text{ref}}$ and $T_{\text{ref}}$ are the reference pressures and temperatures, $R$ is the gas constant and $p$ and $T$ are the temperatures and pressures being extrapolated to. 
To estimate planetesimal mid-mantle viscosities extrapolations using both laws were performed on reference viscosity values for Ganymede \citep{ruckriemen_top-down_2018} (see Table \ref{tab:extrap}) using a range in activation volumes from 2$\times10^{-6}$--2$\times10^{-5}\rm \;m^3mol^{-1}$ \citep{hirth_rheology_2003} and activation energies from 240--570 kJ$\rm mol^{-1}$ \citep{karato_rheology_1993,hirth_rheology_2003}. Assuming constant density core and mantle, planetesimal mid-mantle pressures range from 2\,MPa to 50\,MPa for 100--500\,km bodies.  
\begin{table}[]
    \centering
    \begin{tabular}{|c|c|c|c|}
    \hline
        \etar /Pas & $p_{\text{ref}}$/MPa &  $T_{\text{ref}}$ /K & Extrapolated \etar /Pas  \\\hline
         $10^{19}$--$10^{22}$ & 960  & 1600 & $10^{19}$--$10^{24}$\\
        \hline
    \end{tabular}
    \caption{Reference parameters for viscosity extrapolations for Ganymede \citep{ruckriemen_top-down_2018}. Reference pressure is calculated in the mid-mantle using average values for the Ganymede interior structure model in \citet{ruckriemen_top-down_2018}.}
    \label{tab:extrap}
\end{table}
These extrapolations give possible a wide range \etar from $10^{19}$--$10^{24}$\,Pas due to the large uncertainties in activation volume and activation energy. Equation \ref{eq:arr} and \ref{eq:fk} give values within the same order of magnitude.  

Viscosity has a power law dependence on grain size 
\begin{equation}
    \eta_1 = \eta_2\left(\frac{a_1}{a_2}\right)^m,
\end{equation}
where $a$ is the grain size and $m$ a constant exponent \citep[2 for dislocation creep, 3 for diffusion creep][]{scott_effect_2006}. Grain sizes in planetesimals are likely to be between $10^{-4}$--$10^{-3}$m based on observations from achondrites \citep{hutchison_meteorites_2004,lichtenberg_magma_2019} and grain growth models \citep{monnereau_differentiation_2023}. Extrapolating to these grain sizes from an experimental value of $\eta_2=10^{12.5}$\,Pas for $a_2=10^{-5}$m in the diffusion creep regime \citep{scott_effect_2006} gives $\eta_1 = 10^{15}$--$10^{18}$\,Pas. For an experimental value of $\eta_2=10^{14.5}$\,Pas for $a_2=5\times10^{-5}$m in the dislocation creep regime \citep{scott_effect_2006} gives $\eta_1 = 10^{15}$--$10^{17}$\,Pas. Therefore, \etar=$10^{15}$\,Pas will be used as the minimum value in this work. However, due to the uncertainties in grain size, and differences in lengthscale and stress between experiments and planetesimal interiors these low \etar values ($<10^{18}$\,Pas) are less likely to occur in planetesimals.

\section{Effect of \rcmf on peak temperatures}\label{rcmf}

\begin{figure}
    \centering
    \includegraphics[width=0.7\columnwidth]{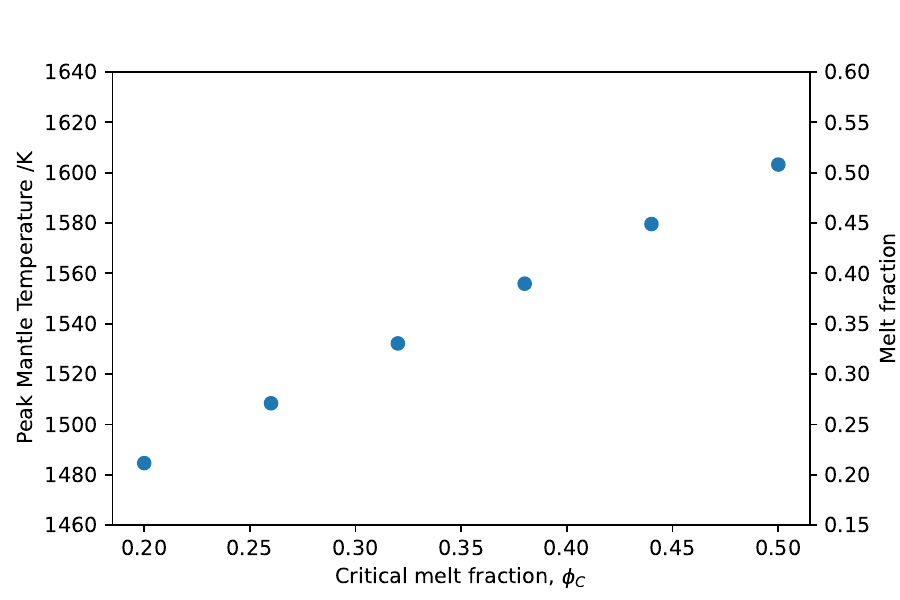}
    \caption{Peak mantle temperature as a function of critical melt fraction, \rcmf.}
    \label{fig:peakT}
\end{figure}

As shown in Figure \ref{fig:peakT}, peak mantle temperature is close to that of \rcmf due to the negative feedback loop between mantle temperature and stagnant lid thickness. At \rcmfp, an increase in temperature leads to a rapid drop in viscosity and thinning of the stagnant lid. This increases surface heat flux, which reduces mantle temperature, increases the viscosity and thickens the stagnant lid. This means choice of \rcmf has direct control on the maximum mantle temperature of a planetesimal.

\bibliographystyle{cas-model2-names}
\bibliography{References}

\begin{thebibliography}{69}
\expandafter\ifx\csname natexlab\endcsname\relax\def\natexlab#1{#1}\fi
\providecommand{\url}[1]{\texttt{#1}}
\providecommand{\href}[2]{#2}
\providecommand{\path}[1]{#1}
\providecommand{\DOIprefix}{doi:}
\providecommand{\ArXivprefix}{arXiv:}
\providecommand{\URLprefix}{URL: }
\providecommand{\Pubmedprefix}{pmid:}
\providecommand{\doi}[1]{\href{http://dx.doi.org/#1}{\path{#1}}}
\providecommand{\Pubmed}[1]{\href{pmid:#1}{\path{#1}}}
\providecommand{\bibinfo}[2]{#2}
\ifx\xfnm\relax \def\xfnm[#1]{\unskip,\space#1}\fi
\bibitem[{Aubert et~al.(2009)Aubert, Labrosse and Poitou}]{aubert_modelling_2009}
\bibinfo{author}{Aubert, J.}, \bibinfo{author}{Labrosse, S.}, \bibinfo{author}{Poitou, C.}, \bibinfo{year}{2009}.
\newblock \bibinfo{title}{Modelling the palaeo-evolution of the geodynamo}.
\newblock \bibinfo{journal}{Geophysical Journal International} \bibinfo{volume}{179}, \bibinfo{pages}{1414--1428}.
\newblock \URLprefix \url{https://academic.oup.com/gji/article/179/3/1414/775893}, \DOIprefix\doi{10.1111/J.1365-246X.2009.04361.X/3/179-3-1414-FIG012.JPEG}. \bibinfo{note}{publisher: Oxford Academic}.
\bibitem[{Borlina et~al.(2022)Borlina, Weiss, Bryson and Armitage}]{borlina_lifetime_2022}
\bibinfo{author}{Borlina, C.S.}, \bibinfo{author}{Weiss, B.P.}, \bibinfo{author}{Bryson, J.F.J.}, \bibinfo{author}{Armitage, P.J.}, \bibinfo{year}{2022}.
\newblock \bibinfo{title}{Lifetime of the {Outer} {Solar} {System} {Nebula} {From} {Carbonaceous} {Chondrites}}.
\newblock \bibinfo{journal}{Journal of Geophysical Research: Planets} \bibinfo{volume}{127}, \bibinfo{pages}{e2021JE007139}.
\newblock \URLprefix \url{https://onlinelibrary.wiley.com/doi/abs/10.1029/2021JE007139}, \DOIprefix\doi{10.1029/2021JE007139}. \bibinfo{note}{\_eprint: https://onlinelibrary.wiley.com/doi/pdf/10.1029/2021JE007139}.
\bibitem[{Borlina et~al.(2021)Borlina, Weiss, Bryson, Bai, Lima, Chatterjee and Mansbach}]{borlina_paleomagnetic_2021}
\bibinfo{author}{Borlina, C.S.}, \bibinfo{author}{Weiss, B.P.}, \bibinfo{author}{Bryson, J.F.J.}, \bibinfo{author}{Bai, X.N.}, \bibinfo{author}{Lima, E.A.}, \bibinfo{author}{Chatterjee, N.}, \bibinfo{author}{Mansbach, E.N.}, \bibinfo{year}{2021}.
\newblock \bibinfo{title}{Paleomagnetic evidence for a disk substructure in the early solar system}.
\newblock \bibinfo{journal}{Science Advances} \bibinfo{volume}{7}, \bibinfo{pages}{eabj6928}.
\newblock \URLprefix \url{https://www.science.org/doi/full/10.1126/sciadv.abj6928}, \DOIprefix\doi{10.1126/sciadv.abj6928}. \bibinfo{note}{publisher: American Association for the Advancement of Science}.
\bibitem[{Bryson et~al.(2019a)Bryson, Neufeld and Nimmo}]{bryson_constraints_2019}
\bibinfo{author}{Bryson, J.F.}, \bibinfo{author}{Neufeld, J.A.}, \bibinfo{author}{Nimmo, F.}, \bibinfo{year}{2019}a.
\newblock \bibinfo{title}{Constraints on asteroid magnetic field evolution and the radii of meteorite parent bodies from thermal modelling}.
\newblock \bibinfo{journal}{Earth and Planetary Science Letters} \bibinfo{volume}{521}, \bibinfo{pages}{68--78}.
\newblock \URLprefix \url{https://doi.org/10.1016/j.epsl.2019.05.046}, \DOIprefix\doi{10.1016/J.EPSL.2019.05.046}. \bibinfo{note}{publisher: Elsevier B.V.}
\bibitem[{Bryson et~al.(2015)Bryson, Nichols, Herrero-Albillos, Kronast, Kasama, Alimadadi, Van Der~Laan, Nimmo and Harrison}]{bryson_long-lived_2015}
\bibinfo{author}{Bryson, J.F.}, \bibinfo{author}{Nichols, C.I.}, \bibinfo{author}{Herrero-Albillos, J.}, \bibinfo{author}{Kronast, F.}, \bibinfo{author}{Kasama, T.}, \bibinfo{author}{Alimadadi, H.}, \bibinfo{author}{Van Der~Laan, G.}, \bibinfo{author}{Nimmo, F.}, \bibinfo{author}{Harrison, R.J.}, \bibinfo{year}{2015}.
\newblock \bibinfo{title}{Long-lived magnetism from solidification-driven convection on the pallasite parent body}.
\newblock \bibinfo{journal}{Nature 2015 517:7535} \bibinfo{volume}{517}, \bibinfo{pages}{472--475}.
\newblock \URLprefix \url{https://www.nature.com/articles/nature14114}, \DOIprefix\doi{10.1038/nature14114}. \bibinfo{note}{publisher: Nature Publishing Group}.
\bibitem[{Bryson et~al.(2023)Bryson, Nichols and Mac~Niocaill}]{bryson_unified_2023}
\bibinfo{author}{Bryson, J.F.J.}, \bibinfo{author}{Nichols, C.I.O.}, \bibinfo{author}{Mac~Niocaill, C.}, \bibinfo{year}{2023}.
\newblock \bibinfo{title}{A unified intensity of the magnetic field in the protoplanetary disk from the {Winchcombe} meteorite}.
\newblock \bibinfo{journal}{Meteoritics \& Planetary Science} \bibinfo{volume}{n/a}, \bibinfo{pages}{1--22}.
\newblock \URLprefix \url{https://onlinelibrary.wiley.com/doi/abs/10.1111/maps.14079}, \DOIprefix\doi{10.1111/maps.14079}. \bibinfo{note}{\_eprint: https://onlinelibrary.wiley.com/doi/pdf/10.1111/maps.14079}.
\bibitem[{Bryson et~al.(2020a)Bryson, Weiss, Biersteker, King and Russell}]{bryson_constraints_2020}
\bibinfo{author}{Bryson, J.F.J.}, \bibinfo{author}{Weiss, B.P.}, \bibinfo{author}{Biersteker, J.B.}, \bibinfo{author}{King, A.J.}, \bibinfo{author}{Russell, S.S.}, \bibinfo{year}{2020}a.
\newblock \bibinfo{title}{Constraints on the {Distances} and {Timescales} of {Solid} {Migration} in the {Early} {Solar} {System} from {Meteorite} {Magnetism}}.
\newblock \bibinfo{journal}{The Astrophysical Journal} \bibinfo{volume}{896}, \bibinfo{pages}{103}.
\newblock \URLprefix \url{https://dx.doi.org/10.3847/1538-4357/ab91ab}, \DOIprefix\doi{10.3847/1538-4357/ab91ab}. \bibinfo{note}{publisher: The American Astronomical Society}.
\bibitem[{Bryson et~al.(2019b)Bryson, Weiss, Getzin, Abrahams, Nimmo and Scholl}]{bryson_paleomagnetic_2019}
\bibinfo{author}{Bryson, J.F.J.}, \bibinfo{author}{Weiss, B.P.}, \bibinfo{author}{Getzin, B.}, \bibinfo{author}{Abrahams, J.N.H.}, \bibinfo{author}{Nimmo, F.}, \bibinfo{author}{Scholl, A.}, \bibinfo{year}{2019}b.
\newblock \bibinfo{title}{Paleomagnetic {Evidence} for a {Partially} {Differentiated} {Ordinary} {Chondrite} {Parent} {Asteroid}}.
\newblock \bibinfo{journal}{Journal of Geophysical Research: Planets} \bibinfo{volume}{124}, \bibinfo{pages}{1880--1898}.
\newblock \URLprefix \url{https://onlinelibrary.wiley.com/doi/abs/10.1029/2019JE005951}, \DOIprefix\doi{10.1029/2019JE005951}. \bibinfo{note}{\_eprint: https://onlinelibrary.wiley.com/doi/pdf/10.1029/2019JE005951}.
\bibitem[{Bryson et~al.(2017)Bryson, Weiss, Harrison, Herrero-Albillos and Kronast}]{bryson_paleomagnetic_2017}
\bibinfo{author}{Bryson, J.F.J.}, \bibinfo{author}{Weiss, B.P.}, \bibinfo{author}{Harrison, R.J.}, \bibinfo{author}{Herrero-Albillos, J.}, \bibinfo{author}{Kronast, F.}, \bibinfo{year}{2017}.
\newblock \bibinfo{title}{Paleomagnetic evidence for dynamo activity driven by inward crystallisation of a metallic asteroid}.
\newblock \bibinfo{journal}{Earth and Planetary Science Letters} \bibinfo{volume}{472}, \bibinfo{pages}{152--163}.
\newblock \URLprefix \url{https://www.sciencedirect.com/science/article/pii/S0012821X17302923}, \DOIprefix\doi{10.1016/j.epsl.2017.05.026}.
\bibitem[{Bryson et~al.(2020b)Bryson, Weiss, Lima, Gattacceca and Cassata}]{bryson_evidence_2020}
\bibinfo{author}{Bryson, J.F.J.}, \bibinfo{author}{Weiss, B.P.}, \bibinfo{author}{Lima, E.A.}, \bibinfo{author}{Gattacceca, J.}, \bibinfo{author}{Cassata, W.S.}, \bibinfo{year}{2020}b.
\newblock \bibinfo{title}{Evidence for {Asteroid} {Scattering} and {Distal} {Solar} {System} {Solids} {From} {Meteorite} {Paleomagnetism}}.
\newblock \bibinfo{journal}{The Astrophysical Journal} \bibinfo{volume}{892}, \bibinfo{pages}{126}.
\newblock \URLprefix \url{https://dx.doi.org/10.3847/1538-4357/ab7cd4}, \DOIprefix\doi{10.3847/1538-4357/ab7cd4}. \bibinfo{note}{publisher: The American Astronomical Society}.
\bibitem[{Carporzen et~al.(2011)Carporzen, Weiss, Elkins-Tanton, Shuster, Ebel and Gattacceca}]{carporzen_magnetic_2011}
\bibinfo{author}{Carporzen, L.}, \bibinfo{author}{Weiss, B.P.}, \bibinfo{author}{Elkins-Tanton, L.T.}, \bibinfo{author}{Shuster, D.L.}, \bibinfo{author}{Ebel, D.}, \bibinfo{author}{Gattacceca, J.}, \bibinfo{year}{2011}.
\newblock \bibinfo{title}{Magnetic evidence for a partially differentiated carbonaceous chondrite parent body}.
\newblock \bibinfo{journal}{Proceedings of the National Academy of Sciences} \bibinfo{volume}{108}, \bibinfo{pages}{6386--6389}.
\newblock \URLprefix \url{https://www.pnas.org/doi/abs/10.1073/pnas.1017165108}, \DOIprefix\doi{10.1073/pnas.1017165108}. \bibinfo{note}{publisher: Proceedings of the National Academy of Sciences}.
\bibitem[{Christensen and Aubert(2006)}]{christensen_scaling_2006}
\bibinfo{author}{Christensen, U.R.}, \bibinfo{author}{Aubert, J.}, \bibinfo{year}{2006}.
\newblock \bibinfo{title}{Scaling properties of convection-driven dynamos in rotating spherical shells and application to planetary magnetic fields}.
\newblock \bibinfo{journal}{Geophysical Journal International} \bibinfo{volume}{166}, \bibinfo{pages}{97--114}.
\newblock \URLprefix \url{https://doi.org/10.1111/j.1365-246X.2006.03009.x}, \DOIprefix\doi{10.1111/j.1365-246X.2006.03009.x}.
\bibitem[{Cook et~al.(2021)Cook, Meyer and Sch{\"o}nb{\"a}chler}]{cook_iron_2021}
\bibinfo{author}{Cook, D.L.}, \bibinfo{author}{Meyer, B.S.}, \bibinfo{author}{Sch{\"o}nb{\"a}chler, M.}, \bibinfo{year}{2021}.
\newblock \bibinfo{title}{Iron and {Nickel} {Isotopes} in {IID} and {IVB} {Iron} {Meteorites}: {Evidence} for {Admixture} of an {SN} {II} {Component} and {Implications} for the {Initial} {Abundance} of {60Fe}}.
\newblock \bibinfo{journal}{The Astrophysical Journal} \bibinfo{volume}{917}, \bibinfo{pages}{59}.
\newblock \URLprefix \url{https://dx.doi.org/10.3847/1538-4357/ac0add}, \DOIprefix\doi{10.3847/1538-4357/ac0add}. \bibinfo{note}{publisher: The American Astronomical Society}.
\bibitem[{Cournede et~al.(2015)Cournede, Gattacceca, Gounelle, Rochette, Weiss and Zanda}]{cournede_early_2015}
\bibinfo{author}{Cournede, C.}, \bibinfo{author}{Gattacceca, J.}, \bibinfo{author}{Gounelle, M.}, \bibinfo{author}{Rochette, P.}, \bibinfo{author}{Weiss, B.P.}, \bibinfo{author}{Zanda, B.}, \bibinfo{year}{2015}.
\newblock \bibinfo{title}{An early solar system magnetic field recorded in {CM} chondrites}.
\newblock \bibinfo{journal}{Earth and Planetary Science Letters} \bibinfo{volume}{410}, \bibinfo{pages}{62--74}.
\newblock \URLprefix \url{https://www.sciencedirect.com/science/article/pii/S0012821X14007110}, \DOIprefix\doi{10.1016/j.epsl.2014.11.019}.
\bibitem[{Courn{\`e}de et~al.(2020)Courn{\`e}de, Gattacceca, Rochette and Shuster}]{cournede_paleomagnetism_2020}
\bibinfo{author}{Courn{\`e}de, C.}, \bibinfo{author}{Gattacceca, J.}, \bibinfo{author}{Rochette, P.}, \bibinfo{author}{Shuster, D.L.}, \bibinfo{year}{2020}.
\newblock \bibinfo{title}{Paleomagnetism of {Rumuruti} chondrites suggests a partially differentiated parent body}.
\newblock \bibinfo{journal}{Earth and Planetary Science Letters} \bibinfo{volume}{533}, \bibinfo{pages}{116042}.
\newblock \URLprefix \url{https://www.sciencedirect.com/science/article/pii/S0012821X1930737X}, \DOIprefix\doi{10.1016/j.epsl.2019.116042}.
\bibitem[{Davies et~al.(2022)Davies, Bono, Meduri, Aubert, Greenwood and Biggin}]{davies_dynamo_2022}
\bibinfo{author}{Davies, C.J.}, \bibinfo{author}{Bono, R.K.}, \bibinfo{author}{Meduri, D.G.}, \bibinfo{author}{Aubert, J.}, \bibinfo{author}{Greenwood, S.}, \bibinfo{author}{Biggin, A.J.}, \bibinfo{year}{2022}.
\newblock \bibinfo{title}{Dynamo constraints on the long-term evolution of {Earth}{\textquoteright}s magnetic field strength}.
\newblock \bibinfo{journal}{Geophysical Journal International} \bibinfo{volume}{228}, \bibinfo{pages}{316--336}.
\newblock \URLprefix \url{https://doi.org/10.1093/gji/ggab342}, \DOIprefix\doi{10.1093/gji/ggab342}.
\bibitem[{Dodds et~al.(2021)Dodds, Bryson, Neufeld and Harrison}]{dodds_thermal_2021}
\bibinfo{author}{Dodds, K.H.}, \bibinfo{author}{Bryson, J.F.}, \bibinfo{author}{Neufeld, J.A.}, \bibinfo{author}{Harrison, R.J.}, \bibinfo{year}{2021}.
\newblock \bibinfo{title}{The {Thermal} {Evolution} of {Planetesimals} {During} {Accretion} and {Differentiation}: {Consequences} for {Dynamo} {Generation} by {Thermally}-{Driven} {Convection}}.
\newblock \bibinfo{journal}{Journal of Geophysical Research: Planets} \bibinfo{volume}{126}.
\newblock \DOIprefix\doi{10.1029/2020JE006704}. \bibinfo{note}{publisher: Blackwell Publishing Ltd}.
\bibitem[{Elkins-Tanton et~al.(2011)Elkins-Tanton, Weiss and Zuber}]{elkins-tanton_chondrites_2011}
\bibinfo{author}{Elkins-Tanton, L.T.}, \bibinfo{author}{Weiss, B.P.}, \bibinfo{author}{Zuber, M.T.}, \bibinfo{year}{2011}.
\newblock \bibinfo{title}{Chondrites as samples of differentiated planetesimals}.
\newblock \bibinfo{journal}{Earth and Planetary Science Letters} \bibinfo{volume}{305}, \bibinfo{pages}{1--10}.
\newblock \URLprefix \url{https://www.sciencedirect.com/science/article/pii/S0012821X11001543}, \DOIprefix\doi{10.1016/j.epsl.2011.03.010}.
\bibitem[{Fang et~al.(2024)Fang, Moynier, Chaussidon, Makhatadze and Villeneuve}]{fang_dating_2024}
\bibinfo{author}{Fang, L.}, \bibinfo{author}{Moynier, F.}, \bibinfo{author}{Chaussidon, M.}, \bibinfo{author}{Makhatadze, G.}, \bibinfo{author}{Villeneuve, J.}, \bibinfo{year}{2024}.
\newblock \bibinfo{title}{Dating metal segregation in asteroids using short-lived $^{\textrm{60}}${Fe}}, \bibinfo{publisher}{GOLDSCHMIDT}.
\newblock \URLprefix \url{https://conf.goldschmidt.info/goldschmidt/2024/meetingapp.cgi/Paper/23986}.
\bibitem[{Fu et~al.(2020)Fu, Kehayias, Weiss, Schrader, Bai and Simon}]{fu_weak_2020}
\bibinfo{author}{Fu, R.R.}, \bibinfo{author}{Kehayias, P.}, \bibinfo{author}{Weiss, B.P.}, \bibinfo{author}{Schrader, D.L.}, \bibinfo{author}{Bai, X.N.}, \bibinfo{author}{Simon, J.B.}, \bibinfo{year}{2020}.
\newblock \bibinfo{title}{Weak {Magnetic} {Fields} in the {Outer} {Solar} {Nebula} {Recorded} in {CR} {Chondrites}}.
\newblock \bibinfo{journal}{Journal of Geophysical Research: Planets} \bibinfo{volume}{125}, \bibinfo{pages}{e2019JE006260}.
\newblock \URLprefix \url{https://onlinelibrary.wiley.com/doi/abs/10.1029/2019JE006260}, \DOIprefix\doi{10.1029/2019JE006260}. \bibinfo{note}{\_eprint: https://onlinelibrary.wiley.com/doi/pdf/10.1029/2019JE006260}.
\bibitem[{Fu et~al.(2014a)Fu, Lima and Weiss}]{fu_no_2014}
\bibinfo{author}{Fu, R.R.}, \bibinfo{author}{Lima, E.A.}, \bibinfo{author}{Weiss, B.P.}, \bibinfo{year}{2014}a.
\newblock \bibinfo{title}{No nebular magnetization in the {Allende} {CV} carbonaceous chondrite}.
\newblock \bibinfo{journal}{Earth and Planetary Science Letters} \bibinfo{volume}{404}, \bibinfo{pages}{54--66}.
\newblock \URLprefix \url{https://www.sciencedirect.com/science/article/pii/S0012821X14004646}, \DOIprefix\doi{10.1016/j.epsl.2014.07.014}.
\bibitem[{Fu et~al.(2021)Fu, Volk, Bilardello, Libourel, Lesur and Dor}]{fu_fine-scale_2021}
\bibinfo{author}{Fu, R.R.}, \bibinfo{author}{Volk, M.W.R.}, \bibinfo{author}{Bilardello, D.}, \bibinfo{author}{Libourel, G.}, \bibinfo{author}{Lesur, G.R.J.}, \bibinfo{author}{Dor, O.B.}, \bibinfo{year}{2021}.
\newblock \bibinfo{title}{The {Fine}-{Scale} {Magnetic} {History} of the {Allende} {Meteorite}: {Implications} for the {Structure} of the {Solar} {Nebula}}.
\newblock \bibinfo{journal}{AGU Advances} \bibinfo{volume}{2}, \bibinfo{pages}{e2021AV000486}.
\newblock \URLprefix \url{https://onlinelibrary.wiley.com/doi/full/10.1029/2021AV000486}, \DOIprefix\doi{10.1029/2021AV000486}. \bibinfo{note}{publisher: John Wiley \& Sons, Ltd ISBN: 10.1029/2021}.
\bibitem[{Fu et~al.(2014b)Fu, Weiss, Lima, Harrison, Bai, Desch, Ebel, Suavet, Wang, Glenn, Le~Sage, Kasama, Walsworth and Kuan}]{fu_solar_2014}
\bibinfo{author}{Fu, R.R.}, \bibinfo{author}{Weiss, B.P.}, \bibinfo{author}{Lima, E.A.}, \bibinfo{author}{Harrison, R.J.}, \bibinfo{author}{Bai, X.N.}, \bibinfo{author}{Desch, S.J.}, \bibinfo{author}{Ebel, D.S.}, \bibinfo{author}{Suavet, C.}, \bibinfo{author}{Wang, H.}, \bibinfo{author}{Glenn, D.}, \bibinfo{author}{Le~Sage, D.}, \bibinfo{author}{Kasama, T.}, \bibinfo{author}{Walsworth, R.L.}, \bibinfo{author}{Kuan, A.T.}, \bibinfo{year}{2014}b.
\newblock \bibinfo{title}{Solar nebula magnetic fields recorded in the {Semarkona} meteorite}.
\newblock \bibinfo{journal}{Science} \bibinfo{volume}{346}, \bibinfo{pages}{1089--1092}.
\newblock \URLprefix \url{https://www.science.org/doi/10.1126/science.1258022}, \DOIprefix\doi{10.1126/science.1258022}. \bibinfo{note}{publisher: American Association for the Advancement of Science}.
\bibitem[{Gattacceca et~al.(2016)Gattacceca, Weiss and Gounelle}]{gattacceca_new_2016}
\bibinfo{author}{Gattacceca, J.}, \bibinfo{author}{Weiss, B.P.}, \bibinfo{author}{Gounelle, M.}, \bibinfo{year}{2016}.
\newblock \bibinfo{title}{New constraints on the magnetic history of the {CV} parent body and the solar nebula from the {Kaba} meteorite}.
\newblock \bibinfo{journal}{Earth and Planetary Science Letters} \bibinfo{volume}{455}, \bibinfo{pages}{166--175}.
\newblock \URLprefix \url{https://www.sciencedirect.com/science/article/pii/S0012821X16304885}, \DOIprefix\doi{10.1016/j.epsl.2016.09.008}.
\bibitem[{Giordano et~al.(2008)Giordano, Russell and Dingwell}]{giordano_viscosity_2008}
\bibinfo{author}{Giordano, D.}, \bibinfo{author}{Russell, J.K.}, \bibinfo{author}{Dingwell, D.B.}, \bibinfo{year}{2008}.
\newblock \bibinfo{title}{Viscosity of magmatic liquids: {A} model}.
\newblock \bibinfo{journal}{Earth and Planetary Science Letters} \bibinfo{volume}{271}, \bibinfo{pages}{123--134}.
\newblock \URLprefix \url{https://www.sciencedirect.com/science/article/pii/S0012821X08002240}, \DOIprefix\doi{10.1016/j.epsl.2008.03.038}.
\bibitem[{Henke et~al.(2013)Henke, Gail, Trieloff and Schwarz}]{henke_thermal_2013}
\bibinfo{author}{Henke, S.}, \bibinfo{author}{Gail, H.P.}, \bibinfo{author}{Trieloff, M.}, \bibinfo{author}{Schwarz, W.H.}, \bibinfo{year}{2013}.
\newblock \bibinfo{title}{Thermal evolution model for the {H} chondrite asteroid-instantaneous formation versus protracted accretion}.
\newblock \bibinfo{journal}{Icarus} \bibinfo{volume}{226}, \bibinfo{pages}{212--228}.
\newblock \URLprefix \url{http://dx.doi.org/10.1016/j.icarus.2013.05.034}, \DOIprefix\doi{10.1016/J.ICARUS.2013.05.034}.
\bibitem[{Hirth and Kohlstedt(2003)}]{hirth_rheology_2003}
\bibinfo{author}{Hirth, G.}, \bibinfo{author}{Kohlstedt, D.}, \bibinfo{year}{2003}.
\newblock \bibinfo{title}{Rheology of the {Upper} {Mantle} and the {Mantle} {Wedge}: {A} {View} from the {Experimentalists}}, in: \bibinfo{booktitle}{Inside the {Subduction} {Factory}}. \bibinfo{publisher}{American Geophysical Union}. volume \bibinfo{volume}{138} of \textit{\bibinfo{series}{Geophysical {Monograph} {Series}}}, pp. \bibinfo{pages}{83--105}.
\newblock \URLprefix \url{10.1029/GM138}. \bibinfo{note}{doi:10.1029/138GM06}.
\bibitem[{Hutchison(2004)}]{hutchison_meteorites_2004}
\bibinfo{author}{Hutchison, R.}, \bibinfo{year}{2004}.
\newblock \bibinfo{title}{Meteorites: a petrologic, chemical and isotopic synthesis}.
\newblock Number~\bibinfo{number}{2} in \bibinfo{series}{Cambridge planetary science}, \bibinfo{publisher}{Cambridge University Press}, \bibinfo{address}{Cambridge (GB)}.
\bibitem[{Karato and Wu(1993)}]{karato_rheology_1993}
\bibinfo{author}{Karato, S.i.}, \bibinfo{author}{Wu, P.}, \bibinfo{year}{1993}.
\newblock \bibinfo{title}{Rheology of the {Upper} {Mantle}: {A} {Synthesis}}.
\newblock \bibinfo{journal}{Science} \bibinfo{volume}{260}, \bibinfo{pages}{771--778}.
\newblock \URLprefix \url{https://www.science.org/doi/10.1126/science.260.5109.771}, \DOIprefix\doi{10.1126/science.260.5109.771}. \bibinfo{note}{publisher: American Association for the Advancement of Science}.
\bibitem[{Kodol{\'a}nyi et~al.(2022a)Kodol{\'a}nyi, Hoppe, Vollmer, Berndt and M{\"u}ller}]{kodolanyi_early_2022}
\bibinfo{author}{Kodol{\'a}nyi, J.}, \bibinfo{author}{Hoppe, P.}, \bibinfo{author}{Vollmer, C.}, \bibinfo{author}{Berndt, J.}, \bibinfo{author}{M{\"u}ller, M.}, \bibinfo{year}{2022}a.
\newblock \bibinfo{title}{The {Early} {Solar} {System} {Abundance} of {Iron}-60: {New} {Constraints} from {Chondritic} {Silicates}}.
\newblock \bibinfo{journal}{The Astrophysical Journal} \bibinfo{volume}{940}, \bibinfo{pages}{95}.
\newblock \URLprefix \url{https://dx.doi.org/10.3847/1538-4357/ac8b85}, \DOIprefix\doi{10.3847/1538-4357/ac8b85}. \bibinfo{note}{publisher: The American Astronomical Society}.
\bibitem[{Kodol{\'a}nyi et~al.(2022b)Kodol{\'a}nyi, Hoppe, Vollmer, Berndt and M{\"u}ller}]{kodolanyi_iron-60_2022}
\bibinfo{author}{Kodol{\'a}nyi, J.}, \bibinfo{author}{Hoppe, P.}, \bibinfo{author}{Vollmer, C.}, \bibinfo{author}{Berndt, J.}, \bibinfo{author}{M{\"u}ller, M.}, \bibinfo{year}{2022}b.
\newblock \bibinfo{title}{Iron-60 in the {Early} {Solar} {System} {Revisited}: {Insights} from {In} {Situ} {Isotope} {Analysis} of {Chondritic} {Troilite}}.
\newblock \bibinfo{journal}{The Astrophysical Journal} \bibinfo{volume}{929}, \bibinfo{pages}{107}.
\newblock \URLprefix \url{https://dx.doi.org/10.3847/1538-4357/ac5910}, \DOIprefix\doi{10.3847/1538-4357/ac5910}. \bibinfo{note}{publisher: The American Astronomical Society}.
\bibitem[{Lesher and Spera(2015)}]{lesher_chapter_2015}
\bibinfo{author}{Lesher, C.E.}, \bibinfo{author}{Spera, F.J.}, \bibinfo{year}{2015}.
\newblock \bibinfo{title}{Chapter 5 - {Thermodynamic} and {Transport} {Properties} of {Silicate} {Melts} and {Magma}}, in: \bibinfo{editor}{Sigurdsson, H.} (Ed.), \bibinfo{booktitle}{The {Encyclopedia} of {Volcanoes} ({Second} {Edition})}. \bibinfo{publisher}{Academic Press}, \bibinfo{address}{Amsterdam}, pp. \bibinfo{pages}{113--141}.
\newblock \URLprefix \url{https://www.sciencedirect.com/science/article/pii/B9780123859389000055}, \DOIprefix\doi{10.1016/B978-0-12-385938-9.00005-5}.
\bibitem[{Lichtenberg et~al.(2019)Lichtenberg, Keller, Katz, Golabek and Gerya}]{lichtenberg_magma_2019}
\bibinfo{author}{Lichtenberg, T.}, \bibinfo{author}{Keller, T.}, \bibinfo{author}{Katz, R.F.}, \bibinfo{author}{Golabek, G.J.}, \bibinfo{author}{Gerya, T.V.}, \bibinfo{year}{2019}.
\newblock \bibinfo{title}{Magma ascent in planetesimals: {Control} by grain size}.
\newblock \bibinfo{journal}{Earth and Planetary Science Letters} \bibinfo{volume}{507}, \bibinfo{pages}{154--165}.
\newblock \DOIprefix\doi{10.1016/J.EPSL.2018.11.034}. \bibinfo{note}{arXiv: 1802.02157 Publisher: Elsevier}.
\bibitem[{Maurel et~al.(2020)Maurel, Bryson, Lyons, Ball, Chopdekar, Scholl, Ciesla, Bottke and Weiss}]{maurel_meteorite_2020}
\bibinfo{author}{Maurel, C.}, \bibinfo{author}{Bryson, J.F.J.}, \bibinfo{author}{Lyons, R.J.}, \bibinfo{author}{Ball, M.R.}, \bibinfo{author}{Chopdekar, R.V.}, \bibinfo{author}{Scholl, A.}, \bibinfo{author}{Ciesla, F.J.}, \bibinfo{author}{Bottke, W.F.}, \bibinfo{author}{Weiss, B.P.}, \bibinfo{year}{2020}.
\newblock \bibinfo{title}{Meteorite evidence for partial differentiation and protracted accretion of planetesimals}.
\newblock \bibinfo{journal}{Science Advances} \bibinfo{volume}{6}, \bibinfo{pages}{eaba1303}.
\newblock \URLprefix \url{https://www.science.org/doi/10.1126/sciadv.aba1303}, \DOIprefix\doi{10.1126/sciadv.aba1303}. \bibinfo{note}{publisher: American Association for the Advancement of Science}.
\bibitem[{Maurel et~al.(2021)Maurel, Bryson, Shah, Chopdekar, T.~Elkins-Tanton, A.~Raymond and Weiss}]{maurel_long-lived_2021}
\bibinfo{author}{Maurel, C.}, \bibinfo{author}{Bryson, J.F.J.}, \bibinfo{author}{Shah, J.}, \bibinfo{author}{Chopdekar, R.V.}, \bibinfo{author}{T.~Elkins-Tanton, L.}, \bibinfo{author}{A.~Raymond, C.}, \bibinfo{author}{Weiss, B.P.}, \bibinfo{year}{2021}.
\newblock \bibinfo{title}{A {Long}-{Lived} {Planetesimal} {Dynamo} {Powered} by {Core} {Crystallization}}.
\newblock \bibinfo{journal}{Geophysical Research Letters} \bibinfo{volume}{48}, \bibinfo{pages}{e2020GL091917}.
\newblock \URLprefix \url{https://onlinelibrary.wiley.com/doi/abs/10.1029/2020GL091917}, \DOIprefix\doi{10.1029/2020GL091917}. \bibinfo{note}{\_eprint: https://onlinelibrary.wiley.com/doi/pdf/10.1029/2020GL091917}.
\bibitem[{Maurel and Gattacceca(2023)}]{maurel_estimating_2023}
\bibinfo{author}{Maurel, C.}, \bibinfo{author}{Gattacceca, J.}, \bibinfo{year}{2023}.
\newblock \bibinfo{title}{Estimating {Paleointensities} {From} {Chemical} {Remanent} {Magnetizations} of {Magnetite} {Using} {Non}-{Heating} {Methods}}.
\newblock \bibinfo{journal}{Journal of Geophysical Research: Planets} \bibinfo{volume}{128}, \bibinfo{pages}{e2023JE007779}.
\newblock \URLprefix \url{https://onlinelibrary.wiley.com/doi/abs/10.1029/2023JE007779}, \DOIprefix\doi{10.1029/2023JE007779}. \bibinfo{note}{\_eprint: https://onlinelibrary.wiley.com/doi/pdf/10.1029/2023JE007779}.
\bibitem[{Maurel and Gattacceca(2024)}]{maurel_4565-my-old_2024}
\bibinfo{author}{Maurel, C.}, \bibinfo{author}{Gattacceca, J.}, \bibinfo{year}{2024}.
\newblock \bibinfo{title}{A 4,565-{My}-old record of the solar nebula field}.
\newblock \bibinfo{journal}{Proceedings of the National Academy of Sciences} \bibinfo{volume}{121}, \bibinfo{pages}{e2312802121}.
\newblock \URLprefix \url{https://www.pnas.org/doi/10.1073/pnas.2312802121}, \DOIprefix\doi{10.1073/pnas.2312802121}. \bibinfo{note}{publisher: Proceedings of the National Academy of Sciences}.
\bibitem[{Monnereau et~al.(2023)Monnereau, Guignard, N{\'e}ri, Toplis and Quitt{\'e}}]{monnereau_differentiation_2023}
\bibinfo{author}{Monnereau, M.}, \bibinfo{author}{Guignard, J.}, \bibinfo{author}{N{\'e}ri, A.}, \bibinfo{author}{Toplis, M.J.}, \bibinfo{author}{Quitt{\'e}, G.}, \bibinfo{year}{2023}.
\newblock \bibinfo{title}{Differentiation time scales of small rocky bodies}.
\newblock \bibinfo{journal}{Icarus} \bibinfo{volume}{390}, \bibinfo{pages}{115294}.
\newblock \URLprefix \url{https://www.sciencedirect.com/science/article/pii/S0019103522003864}, \DOIprefix\doi{10.1016/j.icarus.2022.115294}.
\bibitem[{Moore and Webb(2013)}]{moore_heat-pipe_2013}
\bibinfo{author}{Moore, W.B.}, \bibinfo{author}{Webb, A.A.G.}, \bibinfo{year}{2013}.
\newblock \bibinfo{title}{Heat-pipe {Earth}}.
\newblock \bibinfo{journal}{Nature 2013 501:7468} \bibinfo{volume}{501}, \bibinfo{pages}{501--505}.
\newblock \URLprefix \url{https://www.nature.com/articles/nature12473}, \DOIprefix\doi{10.1038/nature12473}. \bibinfo{note}{publisher: Nature Publishing Group}.
\bibitem[{Neufeld et~al.(2019)Neufeld, Bryson and Nimmo}]{neufeld_top-down_2019}
\bibinfo{author}{Neufeld, J.A.}, \bibinfo{author}{Bryson, J.F.}, \bibinfo{author}{Nimmo, F.}, \bibinfo{year}{2019}.
\newblock \bibinfo{title}{The {Top}-{Down} {Solidification} of {Iron} {Asteroids} {Driving} {Dynamo} {Evolution}}.
\newblock \bibinfo{journal}{Journal of Geophysical Research: Planets} \bibinfo{volume}{124}, \bibinfo{pages}{1331--1356}.
\newblock \URLprefix \url{https://onlinelibrary.wiley.com/doi/full/10.1029/2018JE005900}, \DOIprefix\doi{10.1029/2018JE005900}. \bibinfo{note}{publisher: John Wiley \& Sons, Ltd}.
\bibitem[{Neumann et~al.(2023)Neumann, Luther, Trieloff, Reger and Bouvier}]{neumann_fitting_2023}
\bibinfo{author}{Neumann, W.}, \bibinfo{author}{Luther, R.}, \bibinfo{author}{Trieloff, M.}, \bibinfo{author}{Reger, P.M.}, \bibinfo{author}{Bouvier, A.}, \bibinfo{year}{2023}.
\newblock \bibinfo{title}{Fitting {Thermal} {Evolution} {Models} to the {Chronological} {Record} of {Erg} {Chech} 002 and {Modeling} the {Ejection} {Conditions} of the {Meteorite}}.
\newblock \bibinfo{journal}{The Planetary Science Journal} \bibinfo{volume}{4}, \bibinfo{pages}{196}.
\newblock \URLprefix \url{https://iopscience.iop.org/article/10.3847/PSJ/acf465/meta}, \DOIprefix\doi{10.3847/PSJ/acf465}. \bibinfo{note}{publisher: IOP Publishing}.
\bibitem[{Nichols et~al.(2021)Nichols, Bryson, Cottrell, Fu, Harrison, Herrero-Albillos, Kronast, Tarduno and Weiss}]{nichols_time-resolved_2021}
\bibinfo{author}{Nichols, C.I.}, \bibinfo{author}{Bryson, J.F.}, \bibinfo{author}{Cottrell, R.D.}, \bibinfo{author}{Fu, R.R.}, \bibinfo{author}{Harrison, R.J.}, \bibinfo{author}{Herrero-Albillos, J.}, \bibinfo{author}{Kronast, F.}, \bibinfo{author}{Tarduno, J.A.}, \bibinfo{author}{Weiss, B.P.}, \bibinfo{year}{2021}.
\newblock \bibinfo{title}{A {Time}-{Resolved} {Paleomagnetic} {Record} of {Main} {Group} {Pallasites}: {Evidence} for a {Large}-{Cored}, {Thin}-{Mantled} {Parent} {Body}}.
\newblock \bibinfo{journal}{Journal of Geophysical Research: Planets} \bibinfo{volume}{126}, \bibinfo{pages}{e2021JE006900}.
\newblock \URLprefix \url{https://onlinelibrary.wiley.com/doi/full/10.1029/2021JE006900}, \DOIprefix\doi{10.1029/2021JE006900}. \bibinfo{note}{publisher: John Wiley \& Sons, Ltd ISBN: 10.1029/2021}.
\bibitem[{Nichols et~al.(2016)Nichols, Bryson, Herrero-Albillos, Kronast, Nimmo and Harrison}]{nichols_pallasite_2016}
\bibinfo{author}{Nichols, C.I.}, \bibinfo{author}{Bryson, J.F.}, \bibinfo{author}{Herrero-Albillos, J.}, \bibinfo{author}{Kronast, F.}, \bibinfo{author}{Nimmo, F.}, \bibinfo{author}{Harrison, R.J.}, \bibinfo{year}{2016}.
\newblock \bibinfo{title}{Pallasite paleomagnetism: {Quiescence} of a core dynamo}.
\newblock \bibinfo{journal}{Earth and Planetary Science Letters} \bibinfo{volume}{441}, \bibinfo{pages}{103--112}.
\newblock \DOIprefix\doi{10.1016/J.EPSL.2016.02.037}. \bibinfo{note}{publisher: Elsevier}.
\bibitem[{Nimmo(2009)}]{nimmo_energetics_2009}
\bibinfo{author}{Nimmo, F.}, \bibinfo{year}{2009}.
\newblock \bibinfo{title}{Energetics of asteroid dynamos and the role of compositional convection}.
\newblock \bibinfo{journal}{Geophysical Research Letters} \bibinfo{volume}{36}.
\newblock \URLprefix \url{https://onlinelibrary.wiley.com/doi/full/10.1029/2009GL037997}, \DOIprefix\doi{10.1029/2009GL037997}. \bibinfo{note}{publisher: John Wiley \& Sons, Ltd}.
\bibitem[{Noack and Breuer(2013)}]{noack_first-_2013}
\bibinfo{author}{Noack, L.}, \bibinfo{author}{Breuer, D.}, \bibinfo{year}{2013}.
\newblock \bibinfo{title}{First- and second-order {Frank}-{Kamenetskii} approximation applied to temperature-, pressure- and stress-dependent rheology}.
\newblock \bibinfo{journal}{Geophysical Journal International} \bibinfo{volume}{195}, \bibinfo{pages}{27--46}.
\newblock \URLprefix \url{https://doi.org/10.1093/gji/ggt248}, \DOIprefix\doi{10.1093/gji/ggt248}.
\bibitem[{O{\textquoteright}Brien et~al.(2020)O{\textquoteright}Brien, Tarduno, Anand, Smirnov, Blackman, Carroll-Nellenback and Krot}]{obrien_arrival_2020}
\bibinfo{author}{O{\textquoteright}Brien, T.}, \bibinfo{author}{Tarduno, J.A.}, \bibinfo{author}{Anand, A.}, \bibinfo{author}{Smirnov, A.V.}, \bibinfo{author}{Blackman, E.G.}, \bibinfo{author}{Carroll-Nellenback, J.}, \bibinfo{author}{Krot, A.N.}, \bibinfo{year}{2020}.
\newblock \bibinfo{title}{Arrival and magnetization of carbonaceous chondrites in the asteroid belt before 4562 million years ago}.
\newblock \bibinfo{journal}{Communications Earth \& Environment} \bibinfo{volume}{1}, \bibinfo{pages}{1--7}.
\newblock \URLprefix \url{https://www.nature.com/articles/s43247-020-00055-w}, \DOIprefix\doi{10.1038/s43247-020-00055-w}. \bibinfo{note}{publisher: Nature Publishing Group}.
\bibitem[{Oran et~al.(2018)Oran, Weiss and Cohen}]{oran_were_2018}
\bibinfo{author}{Oran, R.}, \bibinfo{author}{Weiss, B.P.}, \bibinfo{author}{Cohen, O.}, \bibinfo{year}{2018}.
\newblock \bibinfo{title}{Were chondrites magnetized by the early solar wind?}
\newblock \bibinfo{journal}{Earth and Planetary Science Letters} \bibinfo{volume}{492}, \bibinfo{pages}{222--231}.
\newblock \URLprefix \url{https://www.sciencedirect.com/science/article/pii/S0012821X1830075X}, \DOIprefix\doi{10.1016/j.epsl.2018.02.013}.
\bibitem[{Piralla et~al.(2023)Piralla, Villeneuve, Schnuriger, Bekaert and Marrocchi}]{piralla_unified_2023}
\bibinfo{author}{Piralla, M.}, \bibinfo{author}{Villeneuve, J.}, \bibinfo{author}{Schnuriger, N.}, \bibinfo{author}{Bekaert, D.V.}, \bibinfo{author}{Marrocchi, Y.}, \bibinfo{year}{2023}.
\newblock \bibinfo{title}{A unified chronology of dust formation in the early solar system}.
\newblock \bibinfo{journal}{Icarus} \bibinfo{volume}{394}, \bibinfo{pages}{115427}.
\newblock \URLprefix \url{https://www.sciencedirect.com/science/article/pii/S0019103523000040}, \DOIprefix\doi{10.1016/j.icarus.2023.115427}.
\bibitem[{R{\"u}ckriemen et~al.(2015)R{\"u}ckriemen, Breuer and Spohn}]{ruckriemen_fe_2015}
\bibinfo{author}{R{\"u}ckriemen, T.}, \bibinfo{author}{Breuer, D.}, \bibinfo{author}{Spohn, T.}, \bibinfo{year}{2015}.
\newblock \bibinfo{title}{The {Fe} snow regime in {Ganymede}'s core: {A} deep-seated dynamo below a stable snow zone}.
\newblock \bibinfo{journal}{Journal of Geophysical Research: Planets} \bibinfo{volume}{120}, \bibinfo{pages}{1095--1118}.
\newblock \URLprefix \url{https://onlinelibrary.wiley.com/doi/full/10.1002/2014JE004781}, \DOIprefix\doi{10.1002/2014JE004781}. \bibinfo{note}{publisher: John Wiley \& Sons, Ltd}.
\bibitem[{R{\"u}ckriemen et~al.(2018)R{\"u}ckriemen, Breuer and Spohn}]{ruckriemen_top-down_2018}
\bibinfo{author}{R{\"u}ckriemen, T.}, \bibinfo{author}{Breuer, D.}, \bibinfo{author}{Spohn, T.}, \bibinfo{year}{2018}.
\newblock \bibinfo{title}{Top-down freezing in a {Fe}{\textendash}{FeS} core and {Ganymede}{\textquoteright}s present-day magnetic field}.
\newblock \bibinfo{journal}{Icarus} \bibinfo{volume}{307}, \bibinfo{pages}{172--196}.
\newblock \URLprefix \url{https://www.sciencedirect.com/science/article/pii/S0019103517307029}, \DOIprefix\doi{10.1016/j.icarus.2018.02.021}.
\bibitem[{Sanderson(2024)}]{sanderson_refined_2024}
\bibinfo{author}{Sanderson, H.}, \bibinfo{year}{2024}.
\newblock \bibinfo{title}{A refined, versatile model for planetesimal thermal evolution and dynamo generation}.
\newblock \URLprefix \url{https://zenodo.org/records/12771396}, \DOIprefix\doi{10.5281/zenodo.12771396}.
\bibitem[{Sanderson et~al.(2025)Sanderson, Bryson, Nichols and Davies}]{sanderson_unlocking_2025}
\bibinfo{author}{Sanderson, H.R.}, \bibinfo{author}{Bryson, J.F.J.}, \bibinfo{author}{Nichols, C.I.O.}, \bibinfo{author}{Davies, C.J.}, \bibinfo{year}{2025}.
\newblock \bibinfo{title}{Unlocking planetesimal magnetic field histories: {A} refined, versatile model for thermal evolution and dynamo generation}.
\newblock \bibinfo{journal}{Icarus} \bibinfo{volume}{425}, \bibinfo{pages}{116323}.
\newblock \URLprefix \url{https://www.sciencedirect.com/science/article/pii/S001910352400383X}, \DOIprefix\doi{10.1016/j.icarus.2024.116323}.
\bibitem[{Scheinberg et~al.(2016)Scheinberg, Elkins-Tanton, Schubert and Bercovici}]{scheinberg_core_2016}
\bibinfo{author}{Scheinberg, A.}, \bibinfo{author}{Elkins-Tanton, L.T.}, \bibinfo{author}{Schubert, G.}, \bibinfo{author}{Bercovici, D.}, \bibinfo{year}{2016}.
\newblock \bibinfo{title}{Core solidification and dynamo evolution in a mantle-stripped planetesimal}.
\newblock \bibinfo{journal}{Journal of Geophysical Research: Planets} \bibinfo{volume}{121}, \bibinfo{pages}{2--20}.
\newblock \URLprefix \url{https://onlinelibrary.wiley.com/doi/full/10.1002/2015JE004843}, \DOIprefix\doi{10.1002/2015JE004843}. \bibinfo{note}{publisher: John Wiley \& Sons, Ltd}.
\bibitem[{Scott and Kohlstedt(2006)}]{scott_effect_2006}
\bibinfo{author}{Scott, T.}, \bibinfo{author}{Kohlstedt, D.L.}, \bibinfo{year}{2006}.
\newblock \bibinfo{title}{The effect of large melt fraction on the deformation behavior of peridotite}.
\newblock \bibinfo{journal}{Earth and Planetary Science Letters} \bibinfo{volume}{246}, \bibinfo{pages}{177--187}.
\newblock \DOIprefix\doi{10.1016/J.EPSL.2006.04.027}.
\bibitem[{Shah et~al.(2017)Shah, Bates, Muxworthy, Hezel, Russell and Genge}]{shah_long-lived_2017}
\bibinfo{author}{Shah, J.}, \bibinfo{author}{Bates, H.C.}, \bibinfo{author}{Muxworthy, A.R.}, \bibinfo{author}{Hezel, D.C.}, \bibinfo{author}{Russell, S.S.}, \bibinfo{author}{Genge, M.J.}, \bibinfo{year}{2017}.
\newblock \bibinfo{title}{Long-lived magnetism on chondrite parent bodies}.
\newblock \bibinfo{journal}{Earth and Planetary Science Letters} \bibinfo{volume}{475}, \bibinfo{pages}{106--118}.
\newblock \URLprefix \url{https://www.sciencedirect.com/science/article/pii/S0012821X17304193}, \DOIprefix\doi{10.1016/j.epsl.2017.07.035}.
\bibitem[{Sterenborg and Crowley(2012)}]{SterenborgCrowley2013}
\bibinfo{author}{Sterenborg, M.G.}, \bibinfo{author}{Crowley, J.W.}, \bibinfo{year}{2012}.
\newblock \bibinfo{title}{Thermal evolution of early solar system planetesimals and the possibility of sustained dynamos}.
\newblock \bibinfo{journal}{Physics of the Earth and Planetary Interiors} \URLprefix \url{http://dx.doi.org/10.1016/j.pepi.2012.10.006}, \DOIprefix\doi{10.1016/j.pepi.2012.10.006}.
\bibitem[{Stevenson(2003)}]{stevenson_planetary_2003}
\bibinfo{author}{Stevenson, D.J.}, \bibinfo{year}{2003}.
\newblock \bibinfo{title}{Planetary magnetic fields}.
\newblock \bibinfo{journal}{Earth and Planetary Science Letters} \bibinfo{volume}{208}, \bibinfo{pages}{1--11}.
\newblock \URLprefix \url{https://www.sciencedirect.com/science/article/pii/S0012821X02011263}, \DOIprefix\doi{10.1016/S0012-821X(02)01126-3}.
\bibitem[{Sturtz et~al.(2022a)Sturtz, Limare, Chaussidon and Kaminski}]{sturtz_structure_2022}
\bibinfo{author}{Sturtz, C.}, \bibinfo{author}{Limare, A.}, \bibinfo{author}{Chaussidon, M.}, \bibinfo{author}{Kaminski, {\'E}.}, \bibinfo{year}{2022}a.
\newblock \bibinfo{title}{Structure of differentiated planetesimals: {A} chondritic fridge on top of a magma ocean}.
\newblock \bibinfo{journal}{Icarus} \bibinfo{volume}{385}, \bibinfo{pages}{115100}.
\newblock \URLprefix \url{https://www.sciencedirect.com/science/article/pii/S001910352200207X}, \DOIprefix\doi{10.1016/j.icarus.2022.115100}.
\bibitem[{Sturtz et~al.(2022b)Sturtz, Limare, Tait and Kaminski}]{sturtz_birth_2022}
\bibinfo{author}{Sturtz, C.}, \bibinfo{author}{Limare, A.}, \bibinfo{author}{Tait, S.}, \bibinfo{author}{Kaminski, {\'E}.}, \bibinfo{year}{2022}b.
\newblock \bibinfo{title}{Birth and {Decline} of {Magma} {Oceans} in {Planetesimals}: 2. {Structure} and {Thermal} {History} of {Early} {Accreted} {Small} {Planetary} {Bodies}}.
\newblock \bibinfo{journal}{Journal of Geophysical Research: Planets} \bibinfo{volume}{127}, \bibinfo{pages}{e2021JE007020}.
\newblock \URLprefix \url{https://onlinelibrary.wiley.com/doi/abs/10.1029/2021JE007020}, \DOIprefix\doi{10.1029/2021JE007020}. \bibinfo{note}{\_eprint: https://onlinelibrary.wiley.com/doi/pdf/10.1029/2021JE007020}.
\bibitem[{Suttle et~al.(2021)Suttle, King, Schofield, Bates and Russell}]{suttle_aqueous_2021}
\bibinfo{author}{Suttle, M.D.}, \bibinfo{author}{King, A.J.}, \bibinfo{author}{Schofield, P.F.}, \bibinfo{author}{Bates, H.}, \bibinfo{author}{Russell, S.S.}, \bibinfo{year}{2021}.
\newblock \bibinfo{title}{The aqueous alteration of {CM} chondrites, a review}.
\newblock \bibinfo{journal}{Geochimica et Cosmochimica Acta} \bibinfo{volume}{299}, \bibinfo{pages}{219--256}.
\newblock \URLprefix \url{https://www.sciencedirect.com/science/article/pii/S0016703721000363}, \DOIprefix\doi{10.1016/j.gca.2021.01.014}.
\bibitem[{Tang and Dauphas(2012)}]{tang_abundance_2012}
\bibinfo{author}{Tang, H.}, \bibinfo{author}{Dauphas, N.}, \bibinfo{year}{2012}.
\newblock \bibinfo{title}{Abundance, distribution, and origin of {60Fe} in the solar protoplanetary disk}.
\newblock \bibinfo{journal}{Earth and Planetary Science Letters} \bibinfo{volume}{359-360}, \bibinfo{pages}{248--263}.
\newblock \URLprefix \url{https://www.sciencedirect.com/science/article/pii/S0012821X12005705}, \DOIprefix\doi{10.1016/j.epsl.2012.10.011}.
\bibitem[{Tarduno et~al.(2012)Tarduno, Cottrell, Nimmo, Hopkins, Voronov, Erickson, Blackman, Scott and McKinley}]{tarduno_evidence_2012}
\bibinfo{author}{Tarduno, J.A.}, \bibinfo{author}{Cottrell, R.D.}, \bibinfo{author}{Nimmo, F.}, \bibinfo{author}{Hopkins, J.}, \bibinfo{author}{Voronov, J.}, \bibinfo{author}{Erickson, A.}, \bibinfo{author}{Blackman, E.}, \bibinfo{author}{Scott, E.R.}, \bibinfo{author}{McKinley, R.}, \bibinfo{year}{2012}.
\newblock \bibinfo{title}{Evidence for a {Dynamo} in the {Main} {Group} {Pallasite} {Parent} {Body}}.
\newblock \bibinfo{journal}{Science} \bibinfo{volume}{338}, \bibinfo{pages}{939--942}.
\newblock \URLprefix \url{https://www.science.org/doi/full/10.1126/science.1223932}, \DOIprefix\doi{10.1126/science.1223932}. \bibinfo{note}{publisher: American Association for the Advancement of Science}.
\bibitem[{Tarduno et~al.(2017)Tarduno, O'Brien, Blackman and Smirnov}]{tarduno_magnetization_2017}
\bibinfo{author}{Tarduno, J.A.}, \bibinfo{author}{O'Brien, T.M.}, \bibinfo{author}{Blackman, E.G.}, \bibinfo{author}{Smirnov, A.V.}, \bibinfo{year}{2017}.
\newblock \bibinfo{title}{Magnetization of {CV} {Meteorites} in the {Absence} of a {Parent} {Body} {Core} {Dynamo}}, p. \bibinfo{pages}{2850}.
\newblock \URLprefix \url{https://ui.adsabs.harvard.edu/abs/2017LPI....48.2850T}.
\bibitem[{Wang et~al.(2017)Wang, Weiss, Bai, Downey, Wang, Wang, Suavet, Fu and Zucolotto}]{wang_lifetime_2017}
\bibinfo{author}{Wang, H.}, \bibinfo{author}{Weiss, B.P.}, \bibinfo{author}{Bai, X.N.}, \bibinfo{author}{Downey, B.G.}, \bibinfo{author}{Wang, J.}, \bibinfo{author}{Wang, J.}, \bibinfo{author}{Suavet, C.}, \bibinfo{author}{Fu, R.R.}, \bibinfo{author}{Zucolotto, M.E.}, \bibinfo{year}{2017}.
\newblock \bibinfo{title}{Lifetime of the solar nebula constrained by meteorite paleomagnetism}.
\newblock \bibinfo{journal}{Science} \bibinfo{volume}{355}, \bibinfo{pages}{623--627}.
\newblock \URLprefix \url{https://www.science.org/doi/full/10.1126/science.aaf5043}, \DOIprefix\doi{10.1126/science.aaf5043}. \bibinfo{note}{publisher: American Association for the Advancement of Science}.
\bibitem[{Weiss et~al.(2021)Weiss, Bai and Fu}]{weiss_history_2021}
\bibinfo{author}{Weiss, B.P.}, \bibinfo{author}{Bai, X.N.}, \bibinfo{author}{Fu, R.R.}, \bibinfo{year}{2021}.
\newblock \bibinfo{title}{History of the solar nebula from meteorite paleomagnetism}.
\newblock \bibinfo{journal}{Science Advances} \bibinfo{volume}{7}, \bibinfo{pages}{eaba5967}.
\newblock \URLprefix \url{https://www.science.org/doi/full/10.1126/sciadv.aba5967}, \DOIprefix\doi{10.1126/sciadv.aba5967}. \bibinfo{note}{publisher: American Association for the Advancement of Science}.
\bibitem[{Weiss and Elkins-Tanton(2013)}]{weiss_differentiated_2013}
\bibinfo{author}{Weiss, B.P.}, \bibinfo{author}{Elkins-Tanton, L.T.}, \bibinfo{year}{2013}.
\newblock \bibinfo{title}{Differentiated {Planetesimals} and the {Parent} {Bodies} of {Chondrites}}.
\newblock \bibinfo{journal}{Annual Review of Earth and Planetary Sciences} \bibinfo{volume}{41}, \bibinfo{pages}{529--560}.
\newblock \URLprefix \url{https://doi.org/10.1146/annurev-earth-040610-133520}, \DOIprefix\doi{10.1146/annurev-earth-040610-133520}. \bibinfo{note}{\_eprint: https://doi.org/10.1146/annurev-earth-040610-133520}.
\bibitem[{Weiss and Tikoo(2014)}]{weiss_lunar_2014}
\bibinfo{author}{Weiss, B.P.}, \bibinfo{author}{Tikoo, S.M.}, \bibinfo{year}{2014}.
\newblock \bibinfo{title}{The lunar dynamo}.
\newblock \bibinfo{journal}{Science} \bibinfo{volume}{346}.
\newblock \URLprefix \url{https://www.science.org/doi/abs/10.1126/science.1246753}, \DOIprefix\doi{10.1126/SCIENCE.1246753/SUPPL_FILE/WEISS.SM.PDF}. \bibinfo{note}{publisher: American Association for the Advancement of Science}.
\bibitem[{Weiss et~al.(2017)Weiss, Wang, Sharp, Gattacceca, Shuster, Downey, Hu, Fu, Kuan, Suavet, Irving, Wang and Wang}]{weiss_nonmagnetic_2017}
\bibinfo{author}{Weiss, B.P.}, \bibinfo{author}{Wang, H.}, \bibinfo{author}{Sharp, T.G.}, \bibinfo{author}{Gattacceca, J.}, \bibinfo{author}{Shuster, D.L.}, \bibinfo{author}{Downey, B.}, \bibinfo{author}{Hu, J.}, \bibinfo{author}{Fu, R.R.}, \bibinfo{author}{Kuan, A.T.}, \bibinfo{author}{Suavet, C.}, \bibinfo{author}{Irving, A.J.}, \bibinfo{author}{Wang, J.}, \bibinfo{author}{Wang, J.}, \bibinfo{year}{2017}.
\newblock \bibinfo{title}{A nonmagnetic differentiated early planetary body}.
\newblock \bibinfo{journal}{Earth and Planetary Science Letters} \bibinfo{volume}{468}, \bibinfo{pages}{119--132}.
\newblock \URLprefix \url{https://www.sciencedirect.com/science/article/pii/S0012821X17301620}, \DOIprefix\doi{10.1016/j.epsl.2017.03.026}.
\bibitem[{Williams(2009)}]{williams_bottom-up_2009}
\bibinfo{author}{Williams, Q.}, \bibinfo{year}{2009}.
\newblock \bibinfo{title}{Bottom-up versus top-down solidification of the cores of small solar system bodies: {Constraints} on paradoxical cores}.
\newblock \bibinfo{journal}{Earth and Planetary Science Letters} \bibinfo{volume}{284}, \bibinfo{pages}{564--569}.
\newblock \DOIprefix\doi{10.1016/J.EPSL.2009.05.019}. \bibinfo{note}{publisher: Elsevier}.

\end{thebibliography}
\end{document}


\maketitle
\tableofcontents
\listoffigures
\listoftables
\clearpage
\begin{figure}
    \centering
    \includegraphics[width=1\textwidth]{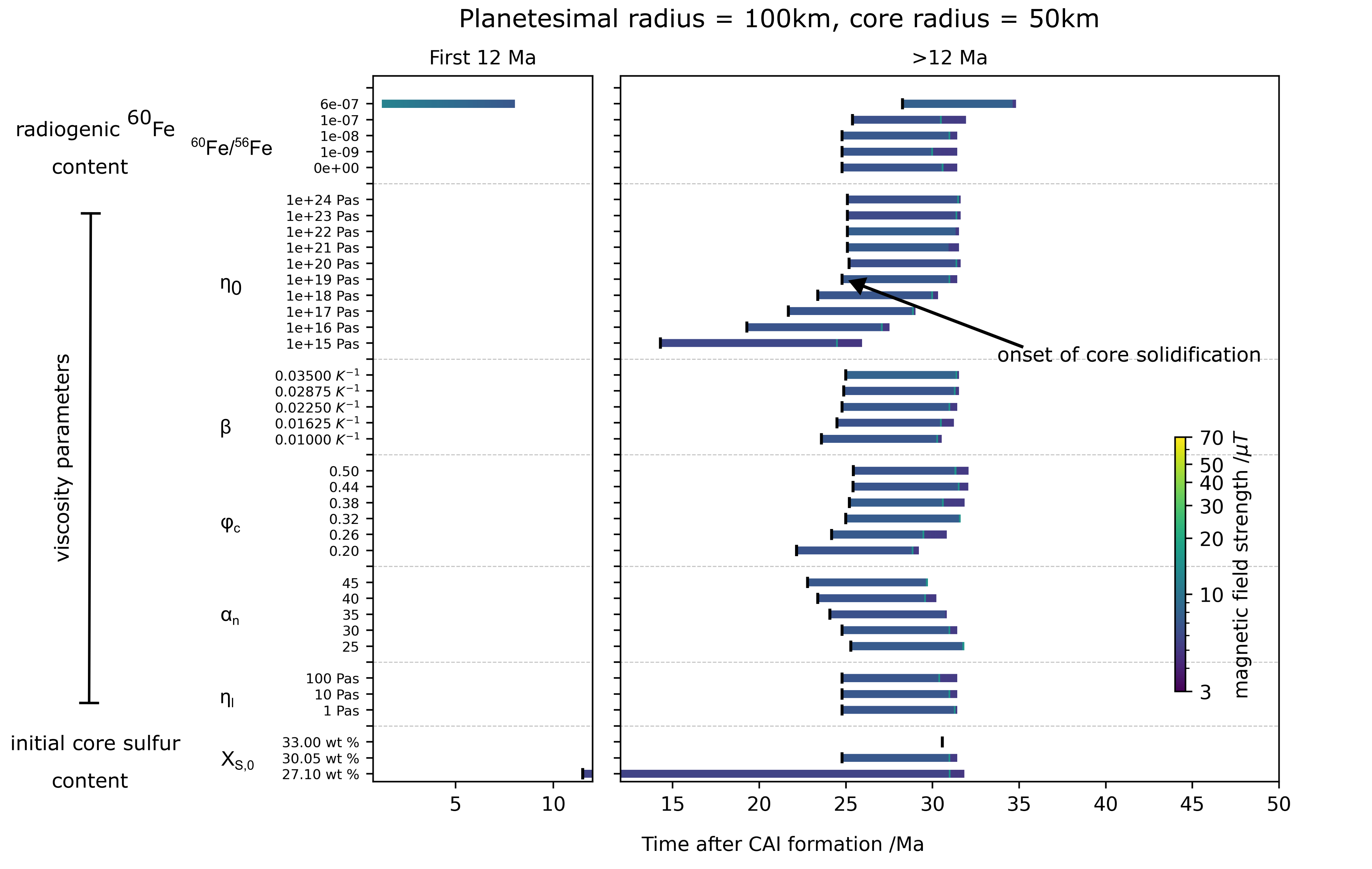}
    \caption[Dynamo timings for a 100\,km radius planetesimal]{Dynamo timings for a 100\,km radius planetesimal for a range of initial \feratiop, mantle viscosities, and core compositions. Filled bars indicate periods when a dynamo is active and the colour of the bar indicates dipole magnetic field strength at the surface. The colour scale is set based on the maximum and minimum field strengths across Figures \ref{fig:dur100}--\ref{fig:dur500}. The small, vertical, black line on each bar indicates the onset of core solidification in each model run. Only the \feratio=$6\times10^{-7}$ run has an epoch of magnetic field generation before the onset of core solidification. For each parameter, the model was run across the range of values in Table 1 in the main text, while all other parameters are held at a constant median value. The constant parameter values  were: $\eta_0=10^{19}$\,Pas, $\beta=0.0225\,K^{-1}$, $\alpha_n=30$, \rcmf= 0.3, $\eta_l=10$\,Pas, \feratio= $10^{-8}$, $X_{S,0}=30.05$\,wt\%. Magnetic field strengths as a function of time for each combination of parameters shown in this plot are provided in the Supplementary Materials.}
    \label{fig:dur100}
\end{figure}

\begin{figure}
    \centering
    \includegraphics[width=1\textwidth]{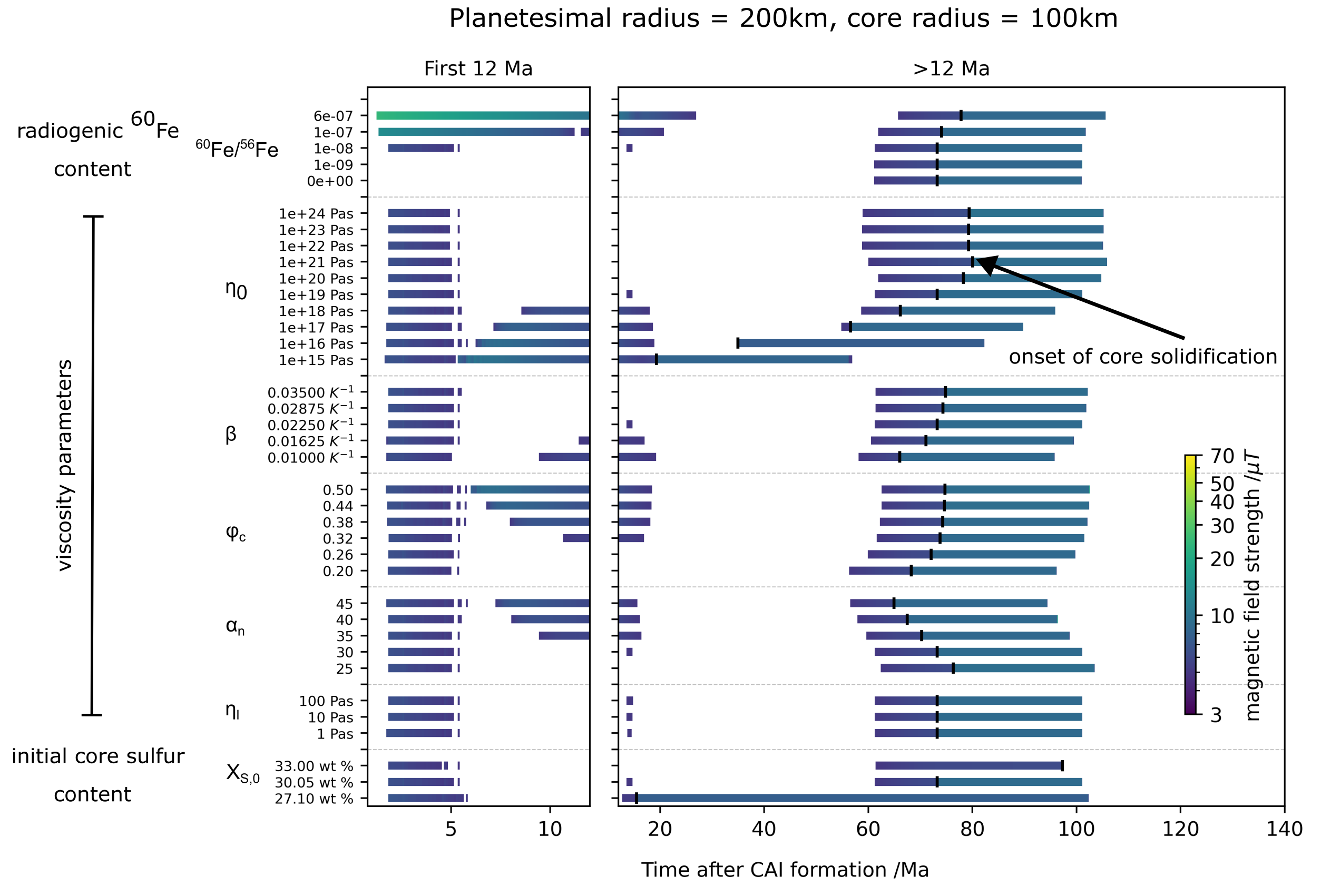}
    \caption[Dynamo timings for a 200\,km planetesimal]{Dynamo timings for a 200\,km radius planetesimal for a range of initial \feratiop, mantle viscosities, and core compositions. Filled bars indicate periods when a dynamo is active and the colour of the bar indicates dipole magnetic field strength at the surface. The colour scale is set based on the maximum and minimum field strengths across Figures \ref{fig:dur100}--\ref{fig:dur500}. The small, vertical, black line on each bar indicates the onset of core solidification in each model run. Magnetic field strengths between the onset of core solidification and the end of dynamo generation are averages for this time interval to remove oscillations in field strength due to the discretisation of the model \citep{sanderson_unlocking_2025}. The thin, vertical, white gaps at early times in some runs are $<0.5$\,Ma breaks in dynamo generation due to a rapid increase in viscosity at \rcmfp. These gaps are artefacts of our viscosity and stagnant lid parametrisation and should not be interpreted. For each parameter, the model was run across the range of values in Table 1 in the main text, while all other parameters are held at a constant median value. The constant parameter values  were: $\eta_0=10^{19}$\,Pas, $\beta=0.0225\,K^{-1}$, $\alpha_n=30$, \rcmf= 0.3, $\eta_l=10$\,Pas, \feratio= $10^{-8}$, $X_{S,0}=30.05$\,wt\%. Magnetic field strengths as a function of time for each combination of parameters shown in this plot are provided in the Supplementary Materials.}
    \label{fig:dur200}
\end{figure}

\begin{figure}
    \centering
    \includegraphics[width=1\textwidth]{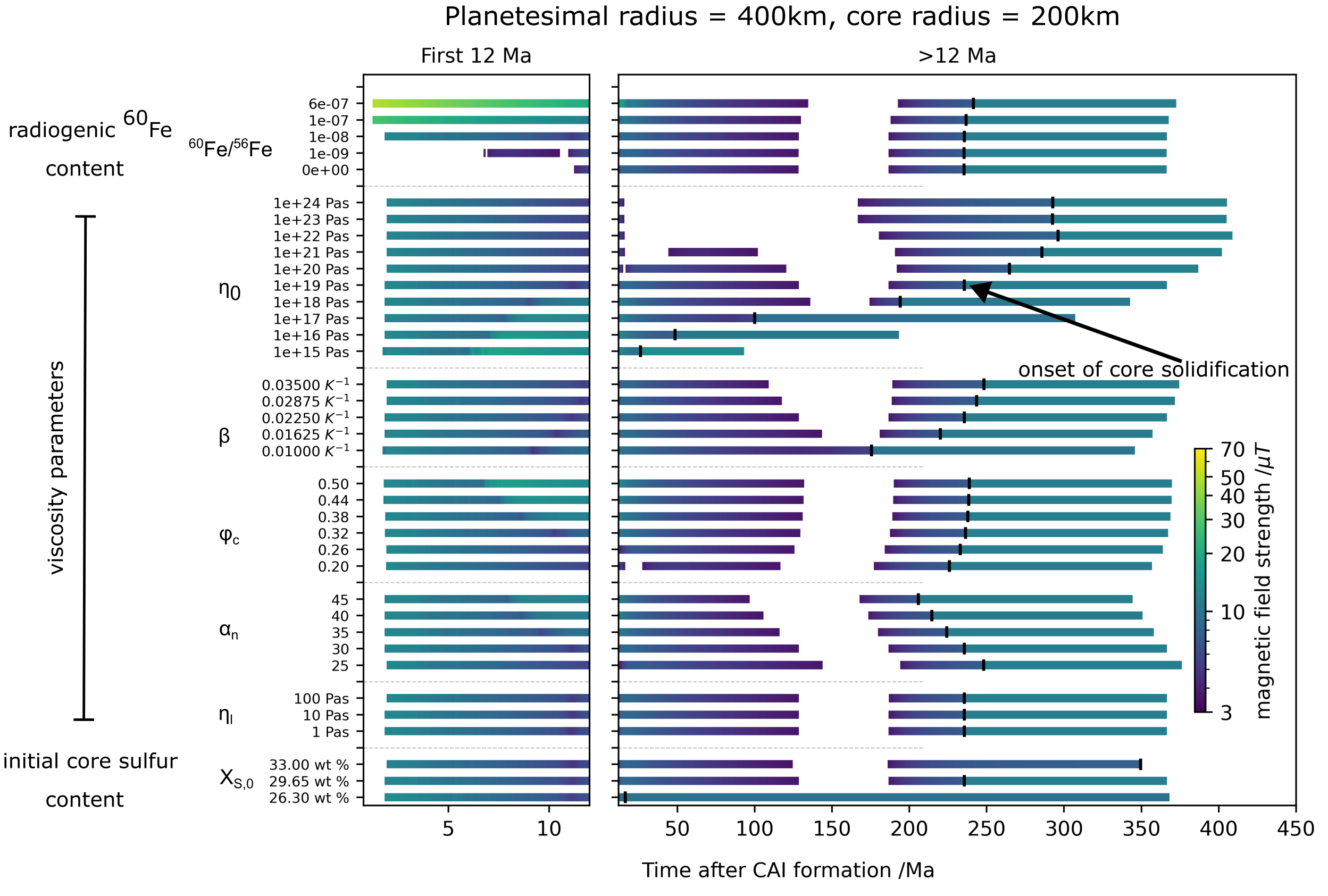}
    \caption[Dynamo timings for a 400\,km radius planetesimal]{Dynamo timings for a 400\,km radius planetesimal for a range of initial \feratiop, mantle viscosities, and core compositions. Filled bars indicate periods when a dynamo is active and the colour of the bar indicates dipole magnetic field strength at the surface. The colour scale is set based on the maximum and minimum field strengths across Figures \ref{fig:dur100}--\ref{fig:dur500}. The small, vertical, black line on each bar indicates the onset of core solidification in each model run. Magnetic field strengths between the onset of core solidification and the end of dynamo generation are averages for this time interval to remove oscillations in field strength due to the discretisation of the model \citep{sanderson_unlocking_2025}. For each parameter, the model was run across the range of values in Table 1 in the main text, while all other parameters are held at a constant median value. The constant parameter values  were: $\eta_0=10^{19}$\,Pas, $\beta=0.0225\,K^{-1}$, $\alpha_n=30$, \rcmf= 0.3, $\eta_l=10$\,Pas, \feratio= $10^{-8}$, $X_{S,0}=29.65$\,wt\%. Magnetic field strengths as a function of time for each combination of parameters shown in this plot are provided in the Supplementary Materials.}
    \label{fig:dur400}
\end{figure}

\begin{figure}
    \centering
    \includegraphics[width=1\textwidth]{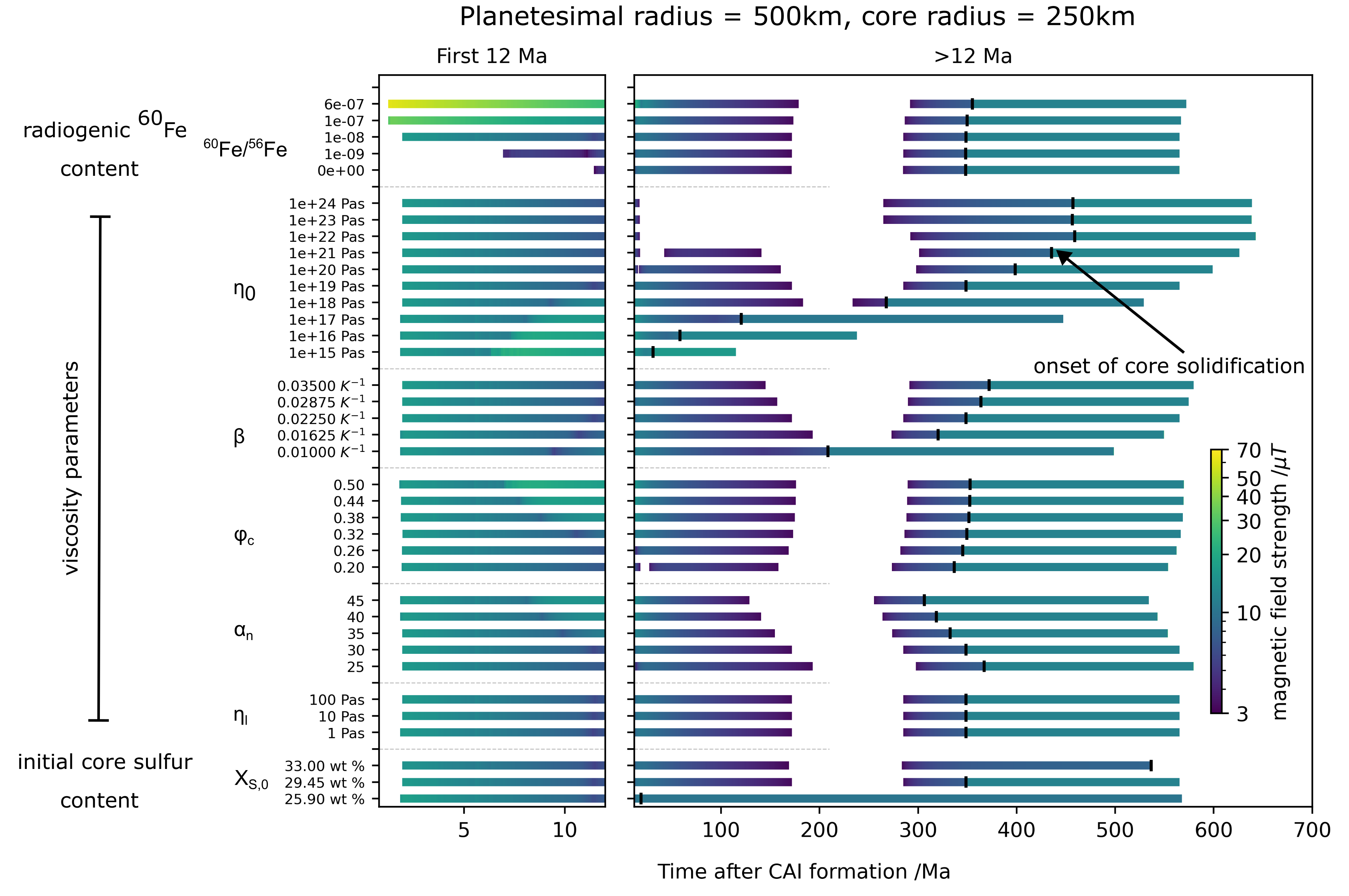}
    \caption[Dynamo timings for a 500\,km radius planetesimal]{Dynamo timings for a 500\,km radius planetesimal for a range of initial \feratiop, mantle viscosities, and core compositions. Filled bars indicate periods when a dynamo is active and the colour of the bar indicates dipole magnetic field strength at the surface. The colour scale is set based on the maximum and minimum field strengths across Figures \ref{fig:dur100}--\ref{fig:dur500}. The small, vertical, black line on each bar indicates the onset of core solidification in each model run. Magnetic field strengths between the onset of core solidification and the end of dynamo generation are averages for this time interval to remove oscillations in field strength due to the discretisation of the model \citep{sanderson_unlocking_2025}. For each parameter, the model was run across the range of values in Table 1 in the main text, while all other parameters are held at a constant median value. The constant parameter values  were: $\eta_0=10^{19}$\,Pas, $\beta=0.0225\,K^{-1}$, $\alpha_n=30$, \rcmf= 0.3, $\eta_l=10$\,Pas, \feratio= $10^{-8}$, $X_{S,0}=29.455$\,wt\%. Magnetic field strengths as a function of time for each combination of parameters shown in this plot are provided in the Supplementary Materials.}
    \label{fig:dur500}
\end{figure}

\section{Range of primordial $^{60}$Fe/$^{56}$Fe}
There are several reasons for the range in the primordial \feratio ratio. First, there is up to two orders of magnitude difference in the values of primordial \feratio between bulk analyses and in-situ studies of individual minerals \citep{kodolanyi_early_2022,kodolanyi_iron-60_2022}. Second, different meteorite groups or minerals produce different values \citep[e.g. $8.8\pm5.2\times10^{-9}$ in angrites vs $7.9\pm3.7\times10^{-7}$ in IID irons;][]{tang_abundance_2012,cook_iron_2021}. Third, biases in error analysis \citep{telus_recalculation_2012} and use of an incorrect reference value for calculating the $^{60}$Ni deficit \citep{kodolanyi_early_2022} can render measurements inaccurate. Fourth, measured systems may not be closed \citep{cook_iron_2021}. Finally, uncertainty in the formation time of a sample increases uncertainty when extrapolating back to the \feratio value at CAI formation \citep[e.g. \feratio$=4\times10^{-8}$--$1.5\times10^{-7}$ for a 5\,Ma range in ages for troilite formation,][]{kodolanyi_iron-60_2022}.

\section{Core solidification endmember}
These model runs use the $m_{frac}=1$ core solidification endmember from \citet{sanderson_unlocking_2025}, because this provides a lower estimate of the magnetic Reynolds number and a more conservative time for when the dynamo will be on. The maximum difference in \Rem between endmembers for a given lengthscale is 16\% \citep{sanderson_unlocking_2025} so the choice of endmember does not have a large effect on the trends observed in this paper.

\section{Angrite paleomagnetism}
\begin{figure}
    \centering
    \includegraphics[width=1\textwidth]{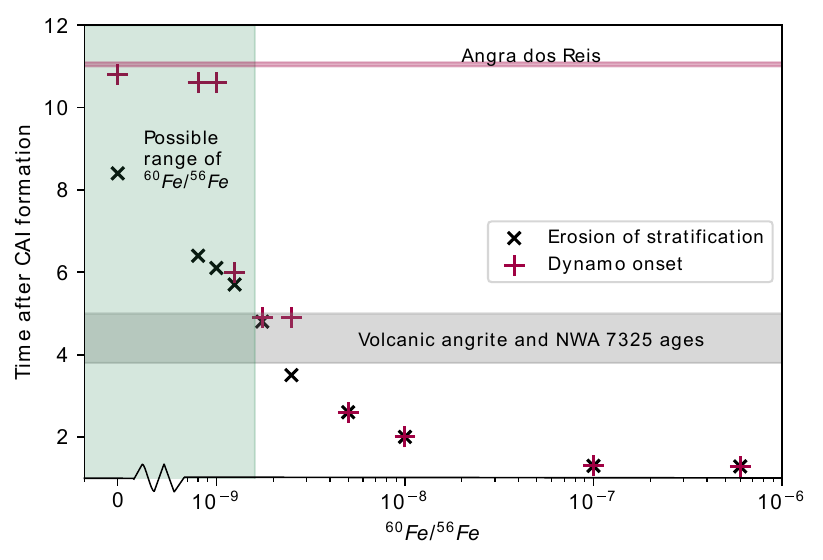}
    \caption[Primordial $^{60}Fe/^{56}Fe$ in angrites]{Primordial $^{60}Fe/^{56}Fe$ required to match the paleomagnetic record of the angrites and NWA 7325. The light green box indicates the range \feratio for which the angrite dynamo starts after the formation of the volcanic angrites and NWA 7325 (grey box), but before the formation of Angra dos Reis (pink box). For the 300\,km body modelled here, \feratio $<1.75\times10^{-9}$ is required to explain these observations. This is a first estimate, which may change with accretion time, body size, and differentiation processes. For \feratio$<10^{-8}$ time taken for erosion of stratification increases linearly as heating to remove the stratification is weaker. There is a lag between dynamo onset time and removal of stratification for \feratio$<1\times10^{-9}$, because as well as lack of heating increasing the time taken to remove stratification it also reduces \fcmbp. Therefore, even once stratification is removed \fcmb may not be high enough to immediately start the dynamo. }
    \label{fig:angrite-fe}
\end{figure}

There are multiple possibilities for the delay in onset of the angrite parent body dynamo relative to our model predictions. From these, a lower \feratio in the formation region of the angrites seems the most likely.
First, the angrite parent body may have been very large \citep[up to Moon-size,][]{tissot_case_2022}, such that our assumptions to neglect the heat of accretion or mantle adiabat may be invalid.
Second, $^{182}$Hf-$^{182}$W measurements suggest angrite core formation occurred in multiple stages, not instantaneous as modelled in this paper \citep{kleine_chronology_2012}. However, gradual addition of material to the core reduces the degree of thermal stratification and enables earlier onset of a dynamo \citep{dodds_thermal_2021}. Additionally, heating by \fe will rapidly remove thermal gradients from gradual accretion. Angrite core formation was complete $\sim2$\,Ma after CAI formation \citep{kleine_chronology_2012}, so late addition of material is unlikely to affect the dynamo a further $\sim$1.8\,Ma after core formation \citep[the time at which the volcanic angrites erupted;][]{wang_lifetime_2017}.
Third, fractionation of iron isotopes during angrite core formation could have altered the proportion of \fe in the core, but this is negligible due to the low pressures in planetesimal size bodies \citep{ni_planet_2022}.
Finally, as discussed in the main text, \feratio may have been lower in the formation region of the angrites. Figure \ref{fig:angrite-fe} displays the range of \feratio values which are consistent with the volcanic and plutonic angrite paleomagnetic record.

\section{Validity of model assumptions} 
A critical magnetic Reynolds number (\Remp) of 10 was used to calculate the dynamo timings in this paper. This is the lowest possible critical value from stability analysis, but numerical simulations for the geodynamo find an empirical critical \Rem of 40--100 \citep{christensen_scaling_2006,stevenson_planetary_2003}. The empirical \Rem $>40$ depends on simulation set-up, such as boundary conditions and inner to outer core ratio, which will differ between planetesimals and the Earth. Figure 4 indicates the times for which the \Rem is greater than each of these critical values. Very few runs reach \Rem $>$40 and only one run for the highest \feratio reaches \Rem $>$100. Furthermore, none of the runs achieve \Rem $>$40 at 100--200\,Ma after CAI formation, where there is paleomagnetic evidence for dynamo generation. This suggests \Rem$>$10 is a reasonable choice for the critical \Rem for planetesimal dynamos.  

Sulfur is assumed to be the only light element in the core. This follows the approach of previous work studying dynamo generation by compositional convection in planetesimals \citep[e.g.][]{nimmo_energetics_2009,scheinberg_core_2016,neufeld_top-down_2019}. Presence of other light elements, such as silicon, oxygen or hydrogen, could affect the core liquidus temperature, the pressure gradient of the liquidus and the density difference during core solidification. This could alter the timing of core solidification and the magnetic field strength generated by a compositional dynamo. However, the liquidus temperature for iron alloyed with these light elements at planetesimal core pressures, and the abundance of these elements in planetesimal cores, is poorly constrained. Therefore, it is not possible to predict in detail how their inclusion in the core could affect these results.

The range of initial core sulfur content, \Xsp, explored in this model is narrow and close to the eutectic composition (26.7--33\,wt\%) due to the constraints posed by differentiation temperature \citep[for more details see][]{sanderson_unlocking_2025}. These sulfur contents are close to those predicted for cores formed from CI, CM, CV, CO, and LL chondrite-like material \citep{bercovici_effects_2022}, so are representative of some planetesimals. Lowering \Xs brings forward the onset of core solidification, but does not appreciably change the end time of dynamo generation. Lower core sulfur contents than explored in this paper could have similar overall dynamo duration, but even earlier onset of core solidification and those bodies may always have thermo-compositional dynamos. Lower \Xs will result in a smaller density difference between the solid iron and liquid Fe-FeS phases, which may reduce the compositional buoyancy flux and dynamo strength.

We assumed that the core radius was half of the planetesimal radius. This is the upper bound of predicted core sizes for cores formed from chondritic starting material \citep[30--50\%][]{bercovici_effects_2022}. However, this core fraction is smaller than that predicted for planetesimals that had their mantles stripped by impacts \citep[e.g. the IVA parent body, asteroid (16) Psyche,][]{yang_iron_2007,elkins-tanton_observations_2020}. We chose a 50\% core radius fraction as a median of these two scenarios. If the core radius was fractionally smaller, the dynamo duration for a given body would be shorter due to a reduced lengthscale for convection and a smaller \Remp. Conversely, a larger core radius fraction would have an increased \Rem and longer dynamo duration. Core solidification would begin earlier in bodies with larger core radius fractions because heat would be lost from the mantle more quickly.

Variable accretion and differentiation via percolation may complicate the effect of \fe on dynamo start time, but heating by \fe should erase differences in core temperature from multiple stages of core formation. Bodies that accrete at the time chosen in this model should reach their critical melt fractions and differentiation via rainout is more applicable than percolation \citep{monnereau_differentiation_2023}. However, future percolation studies should include heating from \fe and explore its effect on differentiation and dynamo generation for later accreting bodies.

\section{Paleomagnetic record data}
See meteorite\_paleomagnetism\_supplementary.csv. 

\section{Effect of viscosity on dynamo timing}
This section provides further detail on the runs with three epochs of dynamo generation.

\subsection{Critical melt fraction, $\phi_C$}
\begin{figure}
    \centering
    \includegraphics[width=1\textwidth]{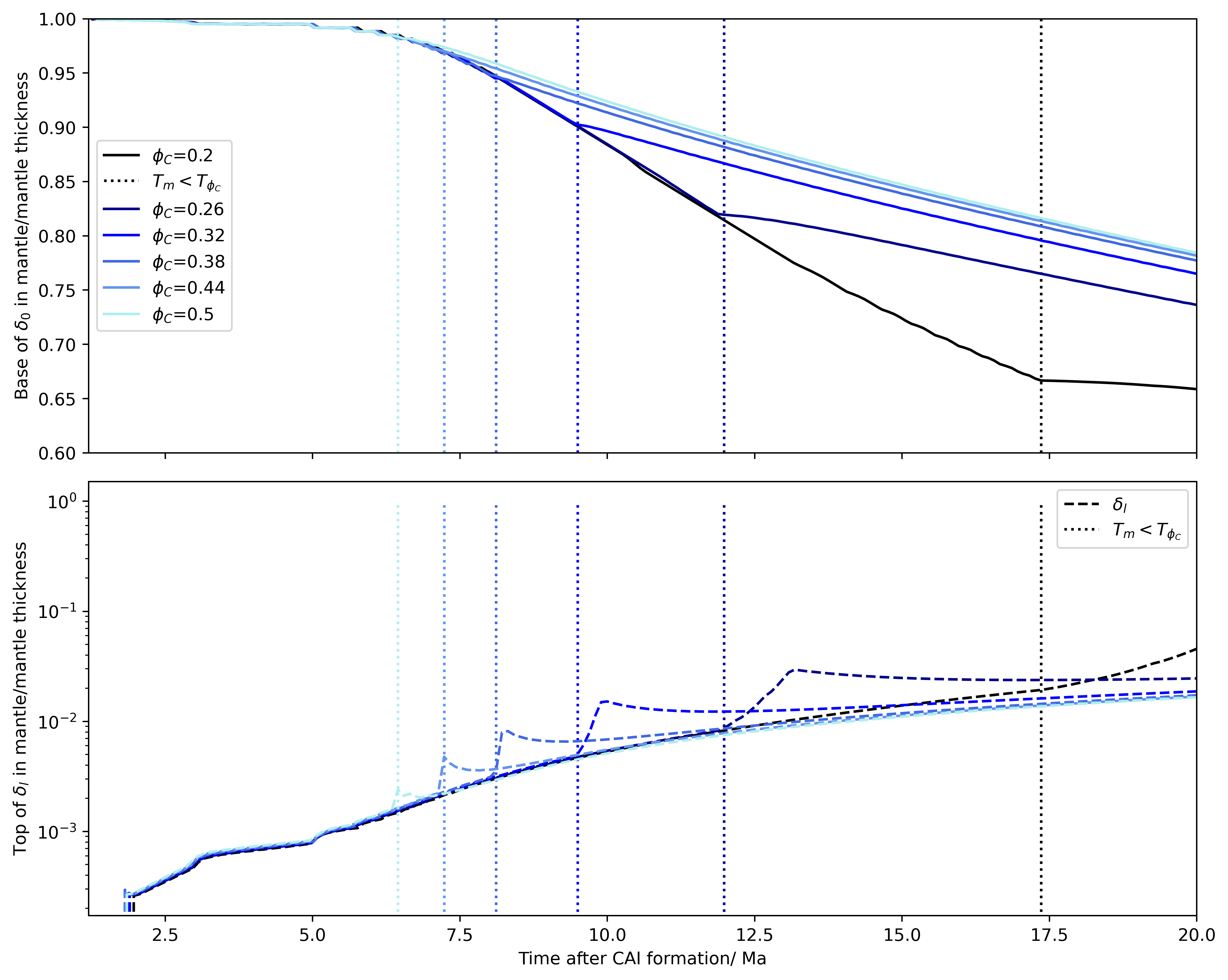}
    \caption[Effect of critical melt fraction, \rcmfp, on boundary layer thicknesses]{Stagnant lid thickness, $\delta_0$, (solid) and CMB boundary layer thickness, $\delta_l$, (dashed) as a function of \rcmf. To the right of the vertical dotted lines indicate when the temperature of the convecting mantle drops below the temperature of \rcmf. This coincides with the uptick in boundary layer thickness as viscosity rapidly increases.}
    \label{fig:supp-rcmf-d0}
\end{figure}

Increasing \rcmf lowers the mantle viscosity at peak temperature, which results in a thinner stagnant lid and longer time before the cessation of convection. In all runs, there is a jump in boundary layer thickness and temporary drop in CMB heat flux when the base of the mantle first cools below the critical melt fraction \citep[see Figure \ref{fig:supp-rcmf-d0} and][]{sanderson_unlocking_2025}. The following increase in CMB heat flux is responsible for the second increase in magnetic field strength \citep{sanderson_unlocking_2025}.

\begin{figure}
    \centering
    \includegraphics[width=1\textwidth]{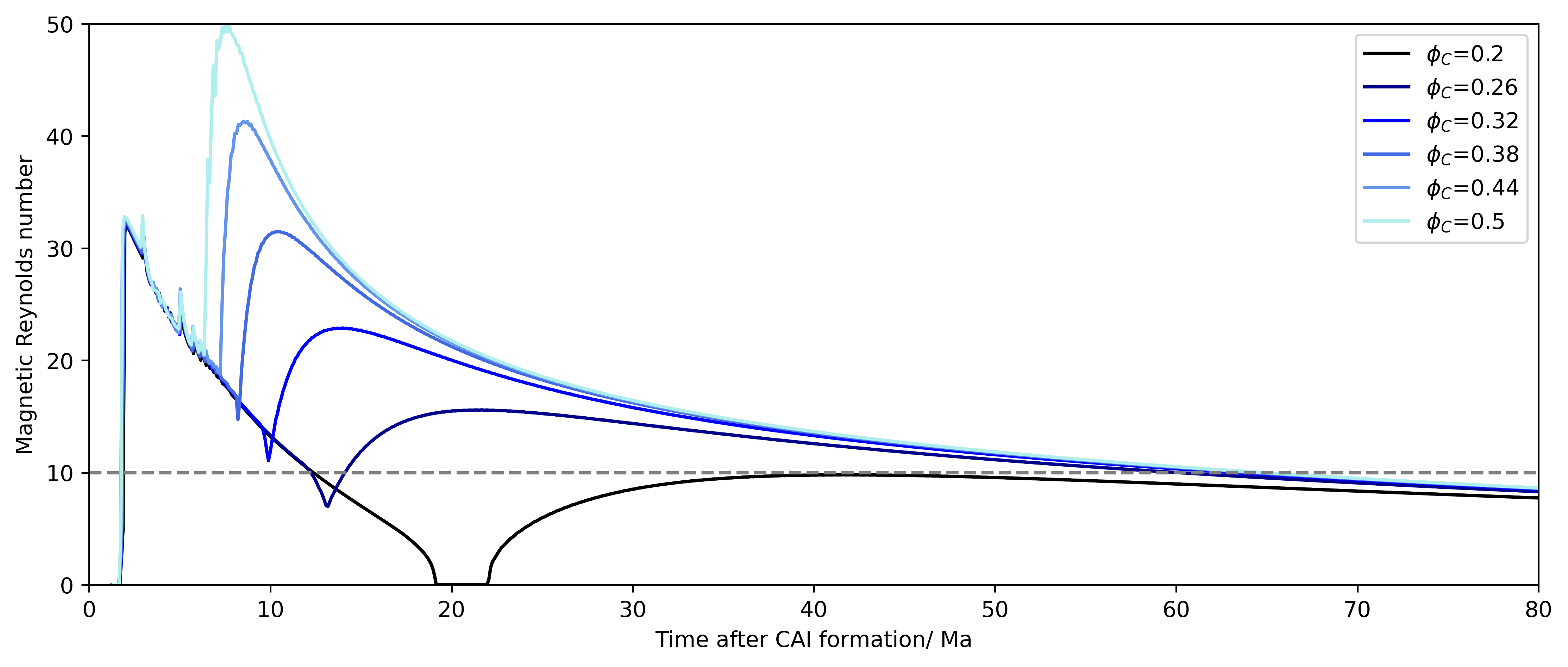}
    \caption[Three epochs of dynamo generation - \rcmf=0.26]{Magnetic Reynolds number, \Remp, as a function of critical melt fraction, \rcmfp. The grey, horizontal, dashed line indicates the critical magnetic Reynolds number. }
    \label{fig:supp-rcmf-rem}
\end{figure}

\begin{figure}
    \centering
    \includegraphics[width=1\textwidth]{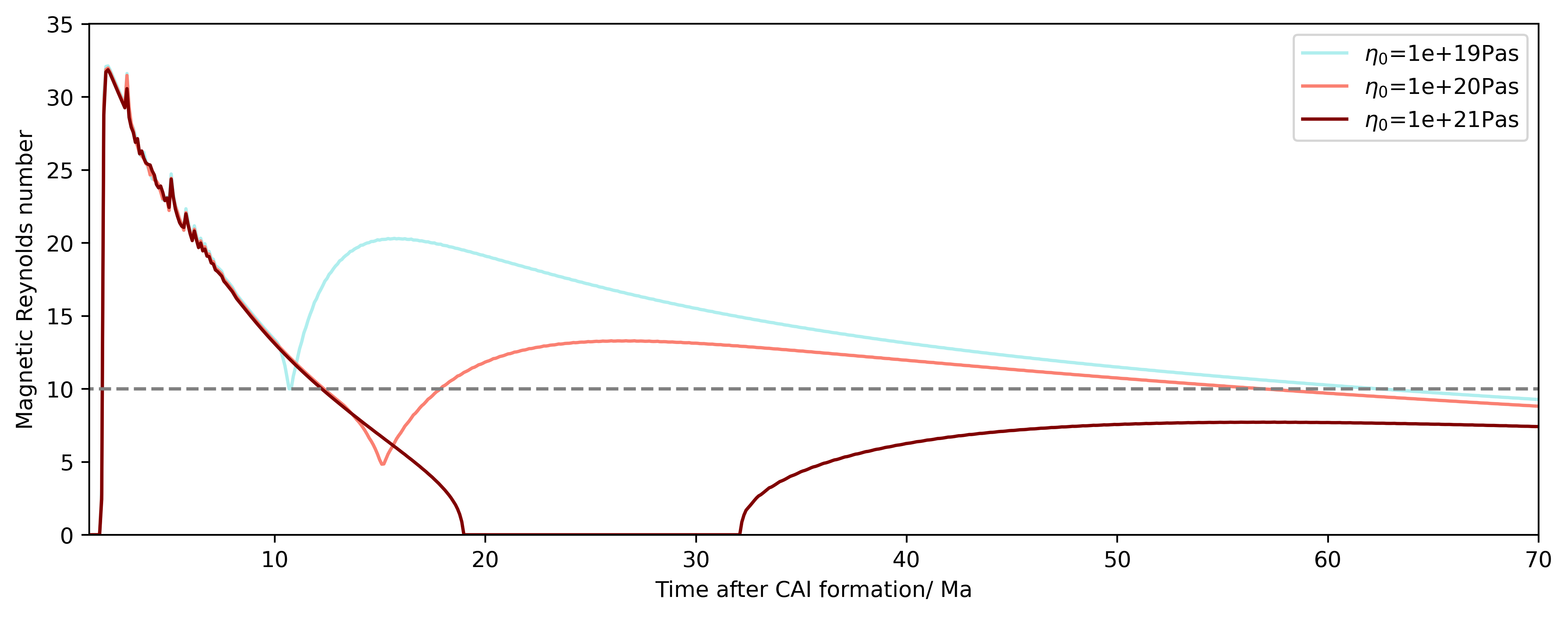}
    \caption[Three epochs of dynamo generation - $\eta_0=10^{20}$\,Pas]{Magnetic Reynolds number, \Remp, for reference viscosities, \etarp, an order of magnitude above and below $10^{20}$\,Pas. The grey, dashed line indicates the critical magnetic Reynolds number.}
    \label{fig:supp-eta0-Rem}
\end{figure}

This jump in CMB boundary layer thickness results in two periods of dynamo generation before the cessation of convection for \rcmf=0.26 and \etar=$10^{20}$\,Pas. In these cases, this jump decreases \Rem enough to temporarily switch off the dynamo (Figure \ref{fig:supp-rcmf-rem} and \ref{fig:supp-eta0-Rem}), but the dynamo restarts after a short period of core cooling. For a more detailed description of CMB boundary layer thickness with time see \citet{sanderson_unlocking_2025}.

\subsection{Reference viscosity, $\eta_0$}
\begin{figure}
    \centering
    \includegraphics[width=1\textwidth]{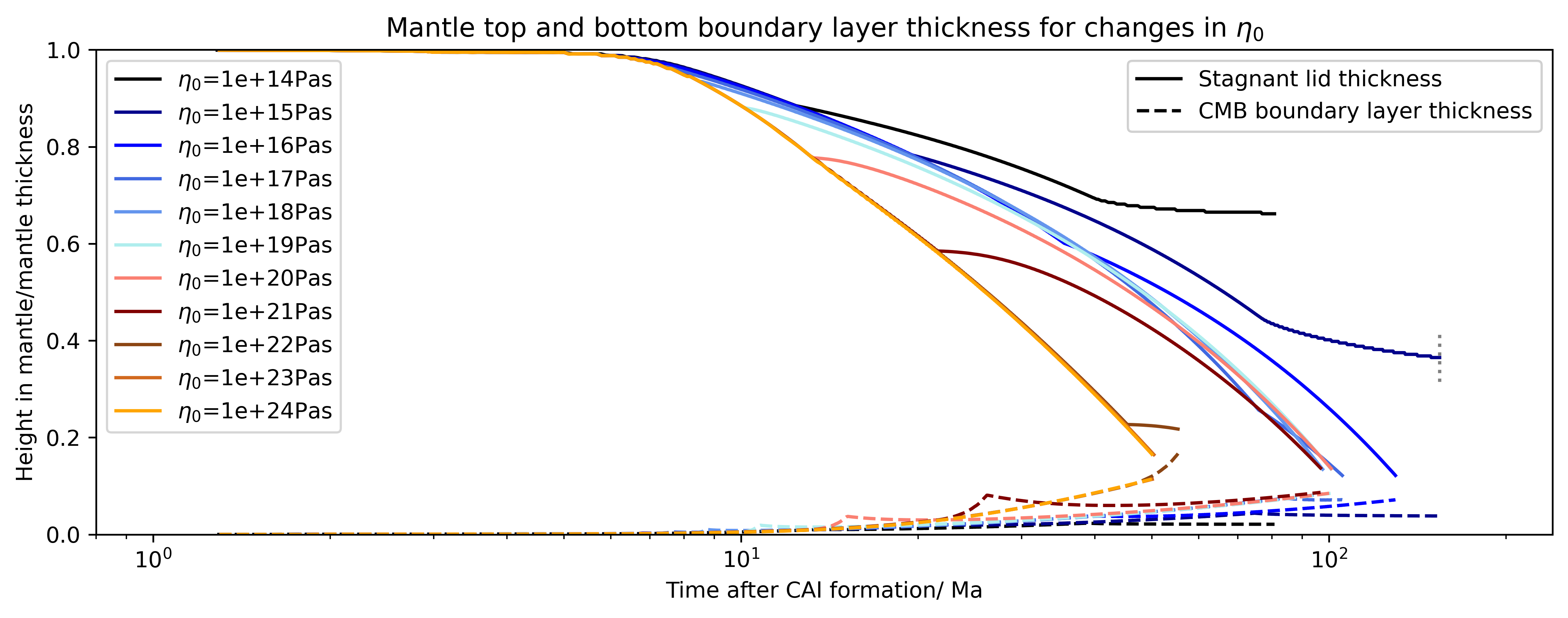}
    \caption[Effect of reference viscosity, \etarp, on boundary layer thickness]{Change in stagnant lid (solid) and CMB boundary layer (dashed) thickness for changes in reference viscosity, \etarp, prior to the cessation of mantle convection. The vertical dotted grey lines on \etar=$10^{14}$,$10^{15}$\,Pas indicate the end of core solidification. Mantle convection ceases when the the stagnant lid and CMB boundary layer meet. Increasing \etar increases the rate at which the stagnant lid thickens, ending mantle convection earlier.}
    \label{fig:supp-eta0-d0}
\end{figure}
For the lowest reference viscosities, \etarp, there is no gap in dynamo generation. This is because heat is removed from the core so rapidly and the mantle has such low viscosity that the core solidifies before the cessation of mantle convection (\etar=$10^{14}$--$10^{15}$\,Pas on Figure \ref{fig:supp-eta0-d0}). 

\subsection{Liquid viscosity, $\eta_l$}
$\eta_l$ has no effect on dynamo timing. This is because the mantle temperature never gets far beyond the critical melt fraction and reaches the fully liquid viscosity due to efficient feedback between stagnant lid thickness and mantle temperature \citep{sanderson_unlocking_2025}. 

\begin{figure}
    \centering
    \includegraphics[width=1\textwidth]{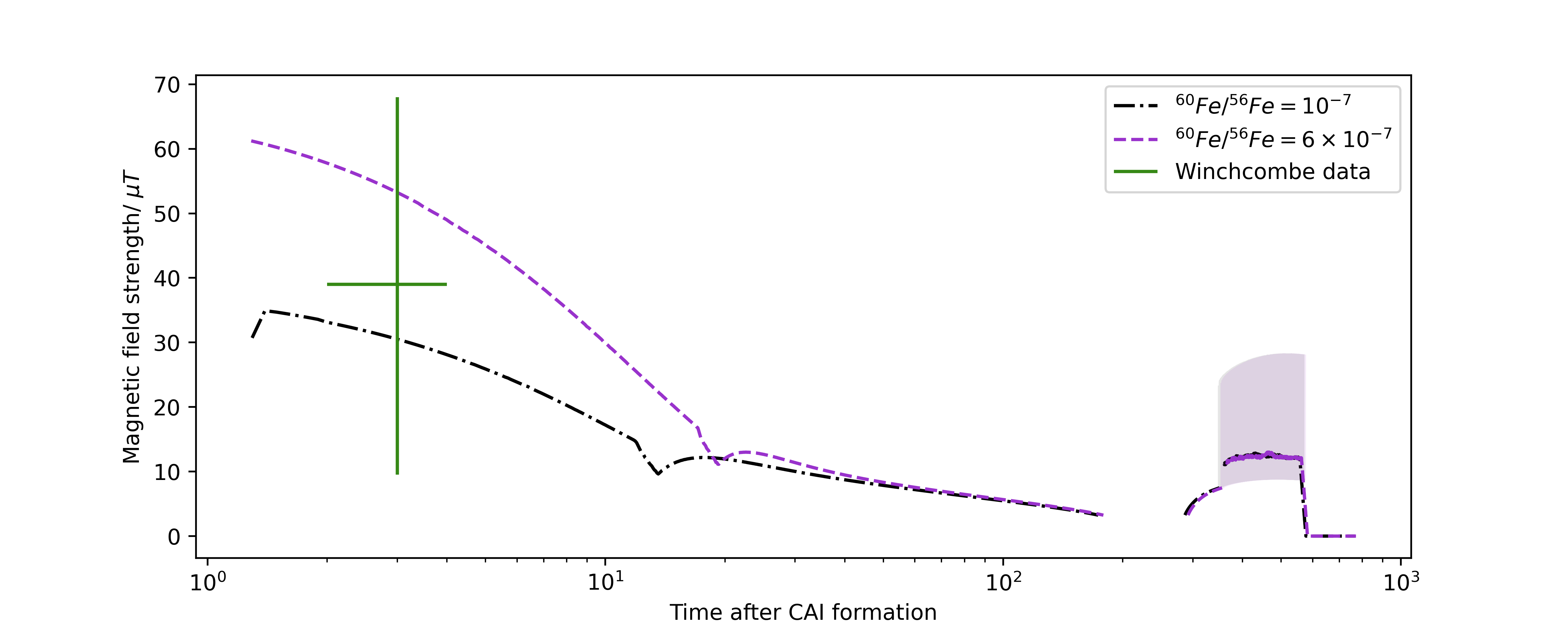}
    \caption[Possible magnetic field histories for the Winchcombe meteorite]{Two example magnetic field histories, which match the intensity (without the factor of two correction) and timing of the paleomagnetic remanance recorded by the Winchcombe meteorite. Both runs are for a 500\,km radius planetesimal, \feratio = $10^{-7}$ or $6\times10^{-7}$ and all other parameters have the constant values in Table 1. Many other parameter combinations may fit the data, this plot aims to illustrate that it is feasible to generate a planetesimal at the correct time and strength to match the data.}
    \label{fig:Winchcombe}
\end{figure}

\clearpage
\bibliography{References}